\documentclass[12pt, a4paper]{article}

\usepackage{url}
\usepackage{amsmath}
\usepackage{amssymb, parskip, listings, mathrsfs, mathtools} 
\usepackage{graphicx}
\usepackage[absolute,overlay]{textpos}
\usepackage{helvet}
\usepackage{array}
\usepackage{color}
\usepackage{setspace}
\usepackage{booktabs}
\usepackage{changepage}
\usepackage{tikz}
\usetikzlibrary{positioning}

\usepackage{geometry, graphicx, amsmath, amssymb, amsthm, parskip, listings, color} 
\usepackage[font=small,labelfont=bf]{caption}
\usepackage{natbib}

\usepackage[affil-it]{authblk} 
\usepackage{etoolbox}
\usepackage{lmodern}



\newcommand*{\affaddr}[1]{#1} 
\newcommand*{\affmark}[1][*]{\textsuperscript{#1}}

\title{\textbf{Bayesian Sparse Mediation Analysis with Targeted Penalization of Natural Indirect Effects}}

\author{
\textbf{Yanyi Song}$^{1}$, \textbf{Xiang Zhou}$^{1,*}$, \textbf{Jian Kang}$^{1,*}$, \textbf{Max T. Aung}$^{1}$, \textbf{Min Zhang}$^{1}$, \textbf{Wei Zhao}$^{2}$, \\ \vspace{-0.1in} \textbf{Belinda L. Needham}$^{2}$, \textbf{Sharon L. R. Kardia}$^{2}$, \textbf{Yongmei Liu}$^{3}$, \textbf{John D. Meeker}$^{4}$,  \textbf{Jennifer A. Smith}$^{2}$, \textbf{and Bhramar Mukherjee}$^{1}$ \\
\affaddr{\affmark[1]Department of Biostatistics, School of Public Health, University of Michigan, Ann Arbor, MI, U.S.A.} \\
\affaddr{\affmark[2]Department of Epidemiology, School of Public Health, University of Michigan, Ann Arbor, MI, U.S.A.}\\
\affaddr{\affmark[3]Department of Medicine, Divisions of Cardiology and Neurology, Duke University, Durham, NC, U.S.A.} \\
\affaddr{\affmark[4]Department of Environmental Health Sciences, School of Public Health, University of Michigan, Ann Arbor, MI, U.S.A.} \\
\affaddr{\affmark[$*$]\textit{email}:\textnormal{ xzhousph@umich.edu}} \\
\affaddr{\affmark[$**$]\textit{email}:\textnormal{ jiankang@umich.edu}}
}

\begin{document}
\date{}
\maketitle

\begin{abstract}
Causal mediation analysis aims to characterize an exposure's effect on an outcome and quantify the indirect effect that acts through a given mediator or a group of mediators of interest. With the increasing availability of measurements on a large number of potential mediators, like the epigenome or the microbiome, new statistical methods are needed to simultaneously accommodate high-dimensional mediators while directly target penalization of the natural indirect effect (NIE) for active mediator identification. Here, we develop two novel prior models for identification of active mediators in high-dimensional mediation analysis through penalizing NIEs in a Bayesian paradigm. Both methods specify a joint prior distribution on the exposure-mediator effect and mediator-outcome effect with either (a) a four-component Gaussian mixture prior or (b) a product threshold Gaussian prior. By jointly modeling the two parameters that contribute to the NIE, the proposed methods enable penalization on their product in a targeted way. Resultant inference can take into account the four-component composite structure underlying the NIE. We show through simulations that the proposed methods improve both selection and estimation accuracy compared to other competing methods. We applied our methods for an in-depth analysis of two ongoing epidemiologic studies: the Multi-Ethnic Study of Atherosclerosis (MESA) and the LIFECODES birth cohort. The identified active mediators in both studies reveal important biological pathways for understanding disease mechanisms.
\end{abstract}

\section{Introduction}
\label{sec1}
Mediation analysis is of increasing importance across a wide range of disciplines (\citealp{mackinnon2007mediation}). It investigates how an intermediate variable, called a mediator, explains the mechanism underlying a known relationship between the exposure and the outcome. The main goal of such an analysis is to disentangle the exposure's effect and identify effect that acts through the mediator of interest, which is often referred to as the indirect/mediation effect. Built on the counterfactual framework, a causal approach to mediation analysis (\citealp{vanderweele2016mediation}) specifies assumptions for a potentially causal interpretation of estimated indirect effects using the classical formulas from \cite{baron1986moderator}. Univariate mediation analysis, which analyzes one mediator at a time, has been studied extensively in areas of social, economic, epidemiological and genetic studies (\citealp{imai2010identification, mackinnon2008introduction}). With the rapid development of high-throughput technologies and the increasing availability of large-scale omics data, however, there is an expanding interest in performing mediation analysis in the presence of a large number of mediators. As one of our motivating examples, the Multi-Ethnic Study of Atherosclerosis (MESA) measured gene expression and DNA methylation (DNAm) levels at the genome-wide scale. These molecular-level omics traits are proposed as part of the mediating mechanism through which neighborhood disadvantages affect physical health (\citealp{smith2017neighborhood}). As another motivating example, the LIFECODES prospective birth cohort collected data on a large group of endogenous biomarkers of lipid metabolism, inflammation, and oxidative stress. These biomarkers are hypothesized to mediate the effects of prenatal exposure to environmental contamination on adverse pregnancy outcomes. Mediation analysis in the above high-dimensional mediator settings can facilitate our understanding of disease etiology but is particularly challenging because the causal ordering among mediators is often unknown {\it a priori} due to a lack of in-depth biological knowledge acquired on the relationship among the mediators. On the other hand, it is not preferable to apply univariate mediation analysis to the high-dimensional setting due to potential confounding of other mediators in the association with the outcome and mis-specification of the true model. 

To enable high-dimensional mediation analysis, several statistical methods have been recently developed. For example, \citet{huang2016hypothesis} and \citet{chen2017high} transform the high-dimensional unordered set of mediators into lower-dimensional orthogonal components using dimension reduction techniques. The extracted low-dimensional components are then analyzed through single mediation analysis. However, it is often not straightforward to interpret the low-dimensional components in these approaches. Shrinkage methods via regularization have also been explored to tackle this high-dimensional regression problem involving two models, the exposure-mediator model and the outcome-exposure model. The Lasso (\citealp{tibshirani1996regression}) penalty can be naturally applied to the two models in mediation analysis. \citet{zhang2016estimating} also proposed a regularized regression with minimax concave penalty for the outcome model after a sure independence screening on mediators. The above methods penalize the mediator-outcome and exposure-mediator coefficients separately without taking into account the structure of the indirect effect. To directly target the mediators with strong indirect effects, \citet{zhao2016pathway} recently developed a new convex, Lasso-type penalty on the indirect effect, which is the product of the two path coefficients. This direct penalization on the pathway effects is shown to improve power for mediator selection and reduce the estimation bias of indirect effects. In addition to frequentist approaches, Bayesian non-parametric models (\citealp{kim2019bayesian}) have been applied in the analysis with a moderate number of mediators. \citet{song2019bayesian} handles high-dimensional mediators through a Bayesian variable selection method and specifies separate shrinkage priors on both the exposure-mediator effects and mediator-outcome effects. However, not modeling the indirect effects in a targeted way may lead to loss of power for selection of active mediators. Therefore, a more effective mediation analysis will require the development of statistical methods that can both handle high-dimensional mediators and select active mediators via direct targeting of the indirect effect.

The indirect effect of a mediator is known to be proportional to the product of the exposure-mediator and mediator-outcome effects under certain assumptions (\citealp{mackinnon2008introduction}). Testing for this product term is not easy due to the complexity in its null distribution. Recent literature began to recognize and leverage the composite structure in the null hypothesis of no indirect effect in the genome-wide mediation analysis setting, where a one-at-a-time single mediator analysis is performed across the entire set of mediators (\citealp{huang2019genome}). Given a large number of mediators, we can characterize the composite space and learn about the structure of mediation through the four components arising from the product of the two effects, i.e. one component of mediators with non-zero indirect effects (active mediators), and three components with zero indirect effects. 

Motivated by the goal of directly targeting the non-null indirect effects to identify active mediators in a high-dimensional mediator setting, we are interested in seeking the Bayesian parallel with a joint prior on the exposure-mediator and mediator-outcome coefficients, which is so far lacking in the literature. One common choice of the bivariate prior would be a Gaussian prior, and it is natural to assume a four-component Gaussian mixture structure on the two effects, corresponding to the composite structure underlying their product. On the other hand, a direct thresholding prior on the indirect effects would also achieve the same goal, and we can extend the hard-thresholding priors (\citealp{ni2019bayesian, cai2020bayesian}) to product thresholding for mediation analysis. Therefore, in this paper, building on the potential outcome framework for causal inference, we develop two novel prior models for high-dimensional mediation analysis: (a) four-component Gaussian mixture prior, and (b) product threshold Gaussian prior. Both models can simultaneously analyze a large number of mediators without making any path-specific or causal ordering assumptions on mediators. The mediator categorization into four groups provides useful interpretations on the way in which a mediator links or does not link exposure to outcome. More importantly, by jointly modeling the exposure-mediator and mediator-outcome coefficients via either bivariate Gaussian distributions or thresholding functions, we place direct shrinkage on the product of the two coefficients, i.e. indirect effect, in a targeted way. Hence, our methods are expected to outperform other penalization methods that apply separate shrinkage in the two regression models independently, for identifying active mediators with non-zero indirect effects.

The proposed methods are generally applicable to many settings, and we examine their performance for both large-scale genomic and environmental data. Specifically, in the MESA cohort, our methods are implemented for high-dimensional mediation analysis with DNAm as mediators (\citealp{bild2002multi}), focusing on the relationship between neighborhood disadvantage (exposure) and body mass index (BMI) as outcome. BMI is a critical risk factor for various diseases like type 2 diabetes (T2D) and cardiovascular disease (CVD) (\citealp{hjellvik2012body}). The important scientific discoveries made in the present study will advance biological understanding of how adverse social circumstances influence our internal molecular environment and in turn lead to cardiometabolic diseases. In the LIFECODES birth cohort, the proposed methods are applied to study the collective impact of endogenous biomarkers in biological pathways in mediating exposure to phthalates (a group of chemicals used to make plastic more flexible) during pregnancy on the gestational age of the newborn at delivery. The integration of molecular/biological data with epidemiologic data in the mediation framework provides interesting and important insights into underlying disease mechanisms. Besides the data analysis, we also perform extensive simulation studies under different structures of effects. We show through both simulations and data analysis that our proposed methods can increase power of a joint analysis and enable efficient identification of individual mediators.

The rest of the paper is organized as follows. In Section \ref{sec2}, we first define the causal effects of interest for the multivariate mediator setting with the counterfactual framework. Then we review the mediation estimands under the regression models with high-dimensional mediators and one continuous outcome. 
In Section \ref{sec4}, we propose two novel methods for direct shrinkage on natural indirect effects. Simulation studies are carried out and discussed in Section \ref{sec5}. We illustrate our methods by  applying them to MESA and LIFECODES data in Section \ref{sec6}, and conclude the paper with a discussion in Section \ref{sec7}.

\section{Notations, Definitions and Models}
\label{sec2}
%
Consider a study of $n$ subjects and for subject $i$, $i=1,\ldots, n$, we collect data on one exposure $A_i$, $p$ candidate mediators $\boldsymbol{M}_i=(M_i^{(1)},M_i^{(2)}, \ldots, M_i^{(p)})^\top$, one outcome $Y_i$, and $q$ covariates $\boldsymbol{C}_i=(C_i^{(1)}, \ldots, C_i^{(q)})^\top$. In particular, we focus on the case where $Y_i$ and $\boldsymbol{M}_i$ are all continuous variables. With the same counterfactual framework as in \citet{song2019bayesian}, let the vector $\boldsymbol{M}_i(a) = (M_i^{(1)}(a),M_i^{(2)}(a), \ldots, M_i^{(p)}(a) )$ denote the $i$th subject's counterfactual value of the $p$ mediators if he/she received exposure $a$. Let $Y_i(a,\boldsymbol{m})$ denote the $i$th subject's counterfactual outcome if the subject's exposure were set to $a$ and mediators were set to $\boldsymbol{m}$. The effect of an exposure can be decomposed into its direct effect and effect mediated through mediators. The natural direct effect (NDE) of the given subject is defined as $Y_i(a, \boldsymbol{M}_i(a^{\star})) - Y_i(a^{\star}, \boldsymbol{M}_i(a^{\star}))$, where the exposure changes from $a^{\star}$ (the reference level) to $a$ and mediators are hypothetically controlled at the level that would have naturally been with exposure $a^{\star}$. The natural indirect effect (NIE) of the given subject is defined by $Y_i(a, \boldsymbol{M}_i(a)) - Y_i(a, \boldsymbol{M}_i(a^{\star}))$. The total effect (TE) can then be expressed as the summation of the NDE and the NIE: $Y_i(a, \boldsymbol{M}_i(a)) - Y_i(a^{\star}, \boldsymbol{M}_i(a^{\star})) =Y_i(a, \boldsymbol{M}_i(a)) - Y_i(a, \boldsymbol{M}_i(a^{\star}))+ Y_i(a, \boldsymbol{M}_i(a^{\star})) - Y_i(a^{\star}, \boldsymbol{M}_i(a^{\star}))  = $ NIE + NDE. 

The counterfactual variables, which are useful concepts to formally define causal effects, are not necessarily observed. To connect the counterfactuals to observed data and estimate the average NDE and NIE from observed data, further assumptions are required, including the consistency assumption and four non-unmeasured confounding assumptions (\citealp{vanderweele2016mediation}). We elaborate those assumptions in Section 1 of the Supplementary Materials (SM). It has been shown that under the required assumptions, the average NDE and NIE can be identified by modeling $Y_i | A_i, \boldsymbol{M}_i, \boldsymbol{C}_i$ and $\boldsymbol{M}_i | A_i, \boldsymbol{C}_i$ using observed data (\citealp{song2019bayesian}). Therefore we can work with the two conditional models for $Y_i | A_i, \boldsymbol{M}_i, \boldsymbol{C}_i$ and $\boldsymbol{M}_i | A_i, \boldsymbol{C}_i$. We propose two linear models for the two conditional relationships $Y_i | A_i, \boldsymbol{M}_i, \boldsymbol{C}_i$ and $\boldsymbol{M}_i | A_i, \boldsymbol{C}_i$. For the outcome model, we assume
\begin{equation}
Y_i = \boldsymbol{M}_i^\top\boldsymbol{\beta_m} + A_i\beta_a + \boldsymbol{C}_i^\top\boldsymbol{\beta_c} + \epsilon_{Yi},
\label{eq:outcome}
\end{equation}
where $\boldsymbol{\beta_m} = (\beta_{m1}, \ldots, \beta_{mp})^\top$; $\boldsymbol{\beta_c} = (\beta_{c1}, \ldots, \beta_{cq})^\top$; and $\epsilon_{Yi} \sim \textnormal{N}(0, \sigma_e^2)$. For the mediator model,  we consider a multivariate regression model that jointly analyzes all $p$ potential mediators together as dependent variables:
\begin{equation}
\boldsymbol{M}_i = A_i\boldsymbol{\alpha_a} + \boldsymbol{\alpha_c}\boldsymbol{C}_i + \boldsymbol{\epsilon}_{Mi},
\label{eq:mediator}
\end{equation}
where $\boldsymbol{\alpha_a} = (\alpha_{a1}, \ldots, \alpha_{ap})^\top$; $\boldsymbol{\alpha_c} = (\boldsymbol{\alpha^\top_{c1}}, \ldots, \boldsymbol{\alpha^\top_{cp}})^\top$; $\boldsymbol{\alpha_{c1}}, \ldots, \boldsymbol{\alpha_{cp}}$ are $q$-by-1 vectors; $\boldsymbol{\epsilon}_{Mi} \sim \mathrm{MVN}(\boldsymbol{0}, \boldsymbol{\Sigma})$, with $\boldsymbol{\Sigma}$ capturing potential residual error covariance. $\epsilon_{Yi}$ and $\boldsymbol{\epsilon}_{Mi}$ are assumed to be independent of each other and independent of $A_i$ and $\boldsymbol{C}_i$. With the identifiability assumptions and the modeling assumptions (linearity, no interaction in the outcome and mediator model) in (\ref{eq:outcome})-(\ref{eq:mediator}), we can compute the average NDE, NIE and TE as below (\citealp{song2019bayesian}). In the rest of the paper, we refer to NDE as direct effect and NIE as indirect/mediation effect.
\begin{eqnarray}
\quad \textnormal{NDE} &=& E[Y_i(a, \boldsymbol{M}_i(a^{\star})) - Y_i(a^{\star}, \boldsymbol{M}_i(a^{\star}))|\boldsymbol{C}_i]=\beta_a(a - a^\star). \\
\quad \textnormal{NIE} &=& E[Y_i(a, \boldsymbol{M}_i(a)) - Y_i(a, \boldsymbol{M}_i(a^{\star}))|\boldsymbol{C}_i] = (a - a^\star)\sum_{j=1}^p \alpha_{aj}\beta_{mj}. \label{eq:sumIE}\\
\quad \textnormal{TE} &=& E[Y_i(a, \boldsymbol{M}_i(a)) - Y_i(a^{\star}, \boldsymbol{M}_i(a^{\star}))|\boldsymbol{C}_i] = (\beta_a + \boldsymbol{\alpha^\top_a}\boldsymbol{\beta_m})(a - a^\star).
\end{eqnarray}
As seen from (\ref{eq:sumIE}), the marginal indirect contribution of the $j$-th mediator is the product of $\alpha_{aj}$ and $\beta_{mj}$. We jointly model $\beta_{mj}$ and $\alpha_{aj}$ and perform targeted shrinkage on the NIE using two prior models described in the next section.

\section{Methods}
\label{sec4}
\subsection{Gaussian Mixture Model (GMM)}
\label{sec41}
The first model we develop to characterize the composite structure of the exposure-mediator and mediator-outcome effects in mediation analysis and induce targeted shrinkage on NIE is the four-component Gaussian mixture model. Mixture models have been studied vastly for classifying subjects into different categories and inferring their association patterns or category-specific properties (\citealp{zeng2018pleiotropic}). In the context of mediation analysis, previous mixture model approaches have primarily been proposed in the form of a principal stratification model (\citealp{gallop2009mediation}).
Here, we introduce a Gaussian mixture model for the joint modeling of $\beta_{mj}$ and $\alpha_{aj}$ and the subsequent inference of the composite association patterns. Specifically, we consider four components in the Gaussian mixture model: a component representing $\beta_{mj}\alpha_{aj} \neq 0$, that both $\beta_{mj}$ and $\alpha_{aj}$ are non-zero; a component representing $\beta_{mj} \neq 0 \textnormal{ and } \alpha_{aj} = 0$; a component representing $\beta_{mj} = 0 \textnormal{ and } \alpha_{aj} \neq 0$; and a component representing $\beta_{mj} = 0 \textnormal{ and } \alpha_{aj} = 0$. To characterize the composite structure underlying the product $\beta_{mj}\alpha_{aj}$, we assume that the effects for each mediator follow a four-component Gaussian mixture distribution as below,
\begin{eqnarray}
[
\beta_{mj},
\alpha_{aj} ]^\top | \{\boldsymbol{V_k}\}_{k=1}^3 \sim \pi_{1}\mathrm{MVN}_2(\boldsymbol{0}, \boldsymbol{V_{1}}) + \pi_{2}\mathrm{MVN}_2(\boldsymbol{0}, \boldsymbol{V_{2}}) \nonumber + \pi_{3}\mathrm{MVN}_2(\boldsymbol{0}, \boldsymbol{V_{3}}) + \pi_{4}\boldsymbol{\delta_0}
\end{eqnarray}
with prior probabilities $\pi_k$ ($k = 1,2,3,4 $) summing to one and $\mathrm{MVN}_2$ denoting a bivariate normal distribution. Here, $\pi_{1}$ represents the prior probability of being an active mediator, with non-zero marginal mediation effect $\beta_{mj}\alpha_{aj}$; and $\boldsymbol{V_{1}}$ models the covariance of $[
\beta_{mj},
\alpha_{aj} ]^\top$ in model (\ref{eq:outcome}) and (\ref{eq:mediator}) when both effects are non-zero. Any inactive mediator will fall into one of the remaining three components. $\pi_{2}$ is the prior probability of having non-zero mediator-outcome effect but zero exposure-mediator effect; and $\boldsymbol{V_{2}} = \begin{bmatrix}
\sigma^2_{2} & 0 \\
0 & 0 \end{bmatrix}$ is a low-rank covariance matrix restricting that only the effect of mediator on outcome $\beta_{mj}$ is non-zero.  $\pi_{3}$ is the prior probability of having non-zero exposure-mediator effect but zero mediator-outcome effect; and $\boldsymbol{V_{3}} = \begin{bmatrix}
0 & 0 \\
0 & \sigma^2_{3} \end{bmatrix}$ is a low-rank covariance matrix restricting that only the effect of exposure on mediator $\alpha_{aj}$ is non-zero. Lastly, $\pi_{4}$ denotes the prior probability of zero mediator-outcome effect and zero exposure-mediator effect; and $\boldsymbol{\delta_0}$ is a point mass at zero. Our method automatically classifies all the mediators into four groups based on their relationship with exposure and outcome. We note that the recently developed Bayesian mediation analysis method (BAMA, \citealp{song2019bayesian}) can be viewed as a two-component version of GMM: in BAMA, the mediator-outcome effect is non-zero and follows a normal distribution with probability $\pi_1+\pi_2$; while the exposure-mediator effect is non-zero and follows another normal distribution with probability $\pi_1+\pi_3$. Consequently, the active mediator in BAMA has \textit{a priori} probability $(\pi_1+\pi_2)(\pi_1+\pi_3)$, which is determined by the non-zero exposure-mediator effect probability and the non-zero mediator-outcome effect probability. 

In GMM, we specify a conjugate inverse-Wishart prior on $\boldsymbol{V_{1}}$, $\boldsymbol{V_{1}} \sim \textnormal{Inv-Wishart}(\boldsymbol{\Psi_0}, 
\nu)$,  where $\boldsymbol{\Psi_0} = \operatorname{diag}\left\{\psi_{01}, \psi_{02}\right\}$ is a diagonal matrix, and $\nu$ is the degree of freedom, and inverse-gamma priors on $\sigma^2_{2}, \sigma^2_{3}$, $\sigma^2_{2} \sim \textnormal{Inv-Gamma}(\nu/2, \psi_{01}/2)$, $\sigma^2_{3} \sim \textnormal{Inv-Gamma}(\nu/2, \psi_{02}/2)$. We also assume $\{ \pi_{1}, \pi_{2}, \pi_{3}, \pi_{4} \} \sim \textnormal{Dirichlet}(a_{1},a_{2},a_{3},a_{4})$
with $a_1$, $a_2$ and $a_3$ set to be smaller than $a_4$ to encourage sparsity of the first three components. For the coefficients of the other covariates, we assume $\beta_a \sim \textnormal{N}(0, \sigma_a^2)$ and  $\boldsymbol{\beta_c}, \boldsymbol{\alpha_{c1}}, ..., \boldsymbol{\alpha_{cp}} \sim \mathrm{MVN}(\boldsymbol{0}, \sigma_c^2 \boldsymbol{I})$. Since we often have adequate information from the data to infer $\boldsymbol{\beta_c}$ and $\boldsymbol{\alpha_c}$, we simply use a limiting prior by setting $\sigma_c^2 \rightarrow \infty$. For the convenience of modeling, we also set the correlation structure among mediators $\boldsymbol{\Sigma}$ as $\sigma_g^2 \boldsymbol{I}$. We use weakly informative inverse-gamma priors on the variance hyper-parameters ($\sigma_a^2, \sigma_e^2$ and $\sigma_g^2 $) in the models.

To facilitate computation, for the $j$th mediator, we create a four-vector membership indicator variable $\boldsymbol{\gamma}_j$, where $\gamma_{jk} = 1$ if $[
\beta_{mj},
\alpha_{aj} ]^\top$ is from normal component $k$ and $\gamma_{jk} = 0$ otherwise. Since the priors used here are all conjugate, we implement a standard Gibbs sampling algorithm and iterate each mediator one at a time to obtain posterior samples. The full details of the algorithm appear in Section 2 of the SM available online. With the Gibbs sampling, for the $j$-th mediator, we can estimate its indirect effect as the product of the posterior mean of $\beta_{mj}$ and $\alpha_{aj}$. We also calculate the posterior probability of both $\beta_{mj}$ and $\alpha_{aj}$ being non-zero as the posterior inclusion probability (PIP), which is $P(\gamma_{j1} = 1|\textnormal{ Data})$. The PIP provides evidence for a non-zero indirect effect, and therefore, we identify mediators with the highest PIP as potentially active mediators.

\subsection{Product Threshold Gaussian (PTG) Prior}
\label{sec42}
Although the GMM model is flexible for a range of applications, the method does not directly impose sparsity on $\beta_{mj}\alpha_{aj}$ for mediator selection. To address this issue, we develop a product threshold Gaussian (PTG) prior for the indirect effects of the $j$-th mediator.
Threshold priors have been recently proposed for Bayesian variable selection. For example,  \citet{ni2019bayesian} introduced a hard-thresholding mechanism in edge selection for sparse graphical structure; \citet{cai2020bayesian} performed a feature selection over networks using the threshold graph Laplacian prior; and \citet{kang2018scalar} developed a soft-thresholding Gaussian process for scalar-on-image regression. As compelling alternatives to shrinkage priors, the threshold priors are equivalent to the non-local priors~(\citealp{rossell2017nonlocal}) which enjoy  appealing theoretical properties and excellent performance in variable selection for high-dimensional regression, especially when the predictors are strongly correlated~(\citealp{kang2018scalar,cai2020bayesian}). In this work, we extend the threshold priors to the product threshold priors for mediation analysis. In particular, for the bivariate vector $(\beta_{mj}, \alpha_{aj})$, $j = 1, ..., p$,
\begin{eqnarray*}
 \beta_{mj} &=& \tilde{\beta}_{mj} \max \left\{I\left(|\tilde{\beta}_{mj}|>\lambda_{1}\right), I\left(|\tilde{\beta}_{mj} \tilde{\alpha}_{aj}|>\lambda_{0}\right)\right\} \\ \alpha_{aj} &=& \tilde{\alpha}_{aj} \max \left\{I\left(|\tilde{\alpha}_{aj}|>\lambda_{2}\right), 
  I\left(|\tilde{\beta}_{mj} \tilde{\alpha}_{aj}|>\lambda_{0}\right)\right\}
\end{eqnarray*}
where the underlying un-thresholded effects $(\tilde{\beta}_{mj}, \tilde{\alpha}_{aj})^\top \sim \mathrm{MVN}_2(0, \boldsymbol{\Sigma_u})$ and $I(\mathcal{A})$ is the indicator function with $I(\mathcal{A}) = 1$ if $\mathcal{A}$ occurs and $I(\mathcal{A}) = 0$ otherwise. We denote $(\beta_{mj}, \alpha_{aj}) \sim \operatorname{PTG}(\boldsymbol{\Sigma_u}, \lambda)$ with $\lambda=\left(\lambda_{0}, \lambda_{1}, \lambda_{2}\right)$ being thresholding parameters.

As one may note, a mediator would escape thresholding and have non-zero indirect effect $\beta_{mj}\alpha_{aj}$ only when (i) both the absolute values of the marginal effects $\tilde{\beta}_{mj}$ and $\tilde{\alpha}_{aj}$ are larger than the threshold values, or (ii) the absolute value of the un-thresholded indirect effect $\tilde{\beta}_{mj}\tilde{\alpha}_{aj}$ is larger than the threshold value. In practice, condition (ii) does not necessarily indicate condition (i). The product threshold prior will facilitate the selection of active mediators by thresholding on the indirect effects in addition to the marginal effects, and shrinking insignificant effects to zero. Similar to GMM, one group of active mediators and three groups of inactive ones are naturally formed. The thresholding on the product term also adds dependency between $\beta_{mj}$ and $\alpha_{aj}$, and we impose no more dependency on the un-thresholded values, namely setting $\boldsymbol{\Sigma_u} = \operatorname{diag}\left\{\tau_{\beta}^{2}, \tau_{\alpha}^{2}\right\}$ in the rest of the paper. 

The threshold parameters $\lambda=\left(\lambda_{0}, \lambda_{1}, \lambda_{2}\right)$ control \textit{a priori} the sparsity of the non-zero effects, and larger values tend to produce a smaller subset of active mediators. Previous literature (\citealp{ni2019bayesian, cai2020bayesian}) have considered uniform priors on those threshold parameters, e.g. $\lambda_0 \sim U[0, \lambda_{0h}]$, $\lambda_1 \sim U[0, \lambda_{1h}]$, $\lambda_2 \sim U[0, \lambda_{2h}]$, with the upper bounds $\lambda_{0h}, \lambda_{1h}, \lambda_{2h}$ being some pre-defined large values. This approach is straightforward and requires little prior knowledge. However, the control of false positives is a concern due to the common under-estimation of $\lambda$. In this paper, we instead determine the threshold parameters from the un-thresholded distributions and the desired number of declared positives, and fix them \textit{a priori}. For example, if we set $\lambda_0 = 0.36, \lambda_1 = \lambda_2 = 0.6$ under $\tau_{\beta}^{2} = 0.1, \tau_{\alpha}^{2} = 0.1$, then the Monte Carlo estimate of the prior proportion of active mediators is approximately 0.01, which could also be tuned to match with $\pi_1$ in the Gaussian mixture model. In practice, we can grid search the three hyper-parameters together with priors on $\tau_{\beta}^{2}$ and $\tau_{\alpha}^{2}$, and find the values that achieve desired prior proportions. The thresholds $\lambda$ can also be interpreted as the minimal detectable signal, and determined based on their practical meaning. Although the resulting selection may be conservative and heavily informed by the pre-defined thresholds, our specification is helpful in guarding against false positive findings. 
As in the GMM model described in \ref{sec41}, conjugate inverse-gamma priors are used for the variance terms ($\tau_{\beta}^{2}, \tau_{\alpha}^{2}, \sigma_e^2$ and $\sigma_g^2$) in the model. The full conditional distributions for $\beta_{mj}$ and $\alpha_{aj}$ are mixtures of truncated normals and can be sampled from Gibbs sampling. The full algorithm appear in Section 3 of the online SM. Similar to GMM, we can calculate the posterior mean of $\beta_{mj}$ and $\alpha_{aj}$, and the posterior probability of both $\beta_{mj}$ and $\alpha_{aj}$ being non-zero as PIP, and use the PIP to rank and select active mediators.

The proposed GMM relies on small values of $\pi_1, \pi_2, \pi_3$ to reflect sparsity on the effects. The Gaussian priors shrink the effects continuously toward zero, and help the model achieve better estimation and prediction performance, but not necessarily mediator selection by the indirect effects. On the other hand, the PTG utilizes a hard threshold function to directly select on the product term $\beta_{mj}\alpha_{aj}$ and map near zero effects to zero. Instead of centering around zero, the effects produced from PTG will be similar to truncated normals away from zero. As a practical procedure, we suggest median inclusion probabilities (PIP = 0.5) as the significance threshold for mediator selection.

\subsection{Other Approaches for High-dimensional Mediation Analysis}
\label{sec43}

Besides GMM and PTG, we also explore a few other approaches. Many of them place simple penalty functions or shrinkage priors on the natural indirect effects.

\noindent\textit{\textbf{Univariate Mediation Analysis}} is perhaps the simplest approach to perform mediation analysis. In univariate mediation analysis, we examine one mediator at a time and test whether the mediator has non-zero indirect effect. We extract \textit{P}-values for testing the indirect effects using the R package \textbf{mediation}.

\noindent\textit{\textbf{Bi-Lasso}} The least absolute shrinkage and selection operator (Lasso) introduced by \cite{tibshirani1996regression} is a widely used penalty function to perform both variable regularization and selection. Here, we consider placing Lasso regularization on the mediator-outcome effects and the exposure-mediator effects separately. For the mediator-outcome effects, we attempt to minimize the following loss function based on the outcome model (\ref{eq:outcome}):
$
f(\boldsymbol{\beta_m}, \beta_a, \boldsymbol{\beta_c}) = \frac{1}{2}\sum_{i=1}^n (Y_i - \boldsymbol{M}_i^\top \boldsymbol{\beta_m} - A_i\beta_a - \boldsymbol{C}_i^\top\boldsymbol{\beta_c} )^2 + \lambda_1\sum_{j=1}^p|\beta_{mj}|
$. For the exposure-mediator effects, we attempt to minimize the following loss function based on the mediator model (\ref{eq:mediator}):
 $
f(\boldsymbol{\alpha_a}, \boldsymbol{\alpha_c}) = \frac{1}{2}\sum_{i=1}^n (\boldsymbol{M}_i -  A_i\boldsymbol{\alpha_a} - \boldsymbol{\alpha_c}\boldsymbol{C}_i)^\top(\boldsymbol{M}_i -  A_i\boldsymbol{\alpha_a}-\boldsymbol{\alpha_c}\boldsymbol{C}_i) + \lambda_2\sum_{j=1}^p|\alpha_{aj}|
$. We perform optimization in the first function using the R package \textbf{glmnet} and perform optimization in the second function using soft-thresholding. We choose the two tuning parameters $\lambda_1 > 0$ and $\lambda_2 > 0$ through 10-fold cross validation in the two functions separately. We refer this approach of applying Lasso separately to the outcome and mediator models as Bi-Lasso.

\noindent\textit{\textbf{Bi-Bayesian Lasso}} is effectively the Bayesian version of Bi-Lasso. It is equivalent to placing a Bayesian Lasso prior (\citealp{park2008bayesian}) on the mediator-outcome effects $\boldsymbol{\beta_m}$ and a separate Bayesian Lasso prior on the exposure-mediator effects $\boldsymbol{\alpha_a}$. Here, we specify the Bayesian Lasso prior for the $j$-th element of $\boldsymbol{\beta_m}$ or $\boldsymbol{\alpha_a}$ as a scale mixture of normal distributions $\textnormal{N}(0, z_j \sigma_{z}^2)$, where the scale parameter $z_j$ follows an exponential distribution $\exp(s^2/2)$ and $1/s^2$ is given a diffuse inverse-gamma prior. We implement the Bi-Bayesian Lasso using a Gibbs sampler following \cite{park2008bayesian} and obtain posterior samples for $\boldsymbol{\beta_m}$ and $\boldsymbol{\alpha_a}$. 

\noindent\textit{\textbf{Pathway Lasso}} is a method developed by \cite{zhao2016pathway} for high-dimensional mediation analysis under the linear structural equation modeling (LSEM) framework. To see how Pathway Lasso works, we first define the squared-error loss in the joint model from equations (\ref{eq:outcome}) and (\ref{eq:mediator}) as
$
l(\boldsymbol{\beta_m}, \boldsymbol{\alpha_a}, \beta_a, \boldsymbol{\beta_c}, \boldsymbol{\alpha_c}) = \sum_{i=1}^n (Y_i - \boldsymbol{M}_i^\top\boldsymbol{\beta_m} - A_i\beta_a - \boldsymbol{C}_i^\top\boldsymbol{\beta_c} )^2 + \sum_{i=1}^n (\boldsymbol{M}_i -  A_i\boldsymbol{\alpha_a} - \boldsymbol{\alpha_c}\boldsymbol{C}_i)^\top(\boldsymbol{M}_i -  A_i\boldsymbol{\alpha_a}-\boldsymbol{\alpha_c}\boldsymbol{C}_i)
$.
The Pathway Lasso then aims to minimize the following penalized function,
\begin{eqnarray}
f(\boldsymbol{\beta_m}, \boldsymbol{\alpha_a}, \beta_a, \boldsymbol{\beta_c}, \boldsymbol{\alpha_c}) &=& \frac{1}{2}l(\boldsymbol{\beta_m}, \boldsymbol{\alpha_a}, \beta_a, \boldsymbol{\beta_c}, \boldsymbol{\alpha_c}) + \lambda[\sum_{j=1}^p\{{|\beta_{mj}\alpha_{aj}|}+ \phi(\beta_{mj}^2+\alpha_{aj}^2)\} + {|\beta_a|}] \nonumber \\ 
&& +  \omega\{\sum_{j=1}^p(|\beta_{mj}|+|\alpha_{aj}|)\}, \phi \geq 1/2  \nonumber \\ 
&=& \frac{1}{2}l(\boldsymbol{\beta_m}, \boldsymbol{\alpha_a}, \beta_a, \boldsymbol{\beta_c}, \boldsymbol{\alpha_c}) + \lambda P_1(\boldsymbol{\beta_m}, \boldsymbol{\alpha_a}, \beta_a) + \omega P_2(\boldsymbol{\beta_m}, \boldsymbol{\alpha_a}) \label{eq:pathwaylasso}
\end{eqnarray}
In Equation (\ref{eq:pathwaylasso}), the first penalty term $P_1$ stabilizes and shrinks the estimates for the indirect effects $\beta_{mj}\alpha_{aj}$. The second penalty term $P_2$ provides additional shrinkage on $\boldsymbol{\beta_m}$ and $\boldsymbol{\alpha_a}$ through a common Lasso penalty placed on both of them. 
We use the algorithm from \cite{zhao2016pathway} to fit Pathway Lasso. We choose the three tuning parameters ($\phi$, $\omega$, and $\lambda$): $\phi = 2$, $\omega = 0.1\lambda$, and choose $\lambda$ through 10-fold cross-validation as in the original paper.

\noindent\textit{\textbf{HIMA}} is another frequentist method developed for high-dimensional mediation analysis (\citealp{zhang2016estimating}). HIMA first applies a sure independence screening to the outcome model to select a small set of potential mediators. With the selected mediators, HIMA then places a minimax concave penalty on the mediator-outcome effects in the outcome model (\ref{eq:outcome}) to obtain effect estimates. The method finally performs a joint significance test and rejects the null hypothesis of no indirect effect with the $j$-th mediator if both $\beta_{mj}$ and $\alpha_{aj}$ are significant. Using the HIMA software, we obtain the Bonferroni corrected \textit{P}-values for testing the indirect effects. 

In addition to the aforementioned methods, we note that several other approaches exist. For example, the methods developed by \citet{huang2016hypothesis} and \citet{chen2017high} first perform dimension reduction on the mediators to extract low dimensional factors on the reduced dimensional space, and then carry out mediation analysis by treating the low dimensional factors as new mediators. Because these approaches analyze the latent factors instead of the original mediators, we do not compare our methods with them in the present study. 

\section{Simulation}
\label{sec5}

\noindent\textit{\textbf{Simulation Overview and Evaluation Metrics}} We evaluate the performance of the two proposed methods (GMM and PTG) and compare them with existing methods in different simulation scenarios. As described in Section \ref{sec4}, we consider a total of eight methods: one univariate method and seven multivariate methods that include four Bayesian methods (GMM, PTG, BAMA and Bi-Bayesian Lasso) and three frequentist methods (Bi-Lasso, Pathway Lasso, and HIMA). We examine the power of different methods to detect true mediators in the simulations. To do so, we rely on PIP to prioritize mediators in PTG, GMM and BAMA; rely on $P$-value to rank mediators in the univariate method and HIMA; and rely on the estimated indirect effects as an measure of evidence for mediation for the remaining methods. To evaluate selection accuracy, we calculate the true positive rate (TPR) based on a fixed false discovery rate (FDR) of 10\% and area under the ROC curve (AUC). To evaluate estimation accuracy, we compute the mean square error (MSE) for the indirect effects of the truly active mediators (MSE$_{\textnormal{non-null}}$), and MSE for the indirect effects of the truly inactive mediators (MSE$_{\textnormal{null}}$). We perform 200 simulation replicates for each scenario to report the average of the above metrics.

\noindent\textit{\textbf{Simulation Design} - Fixed Effect Simulations} We consider one small sample scenario with $n = 100$, $p = 200$, and one large sample scenario with $n = 1000$, $p = 2000$. In both scenarios, we set the proportions of the four different mediator groups to be $\pi_{1} = 0.05, \pi_{2} = 0.05, \pi_{3} = 0.10, \pi_{4} = 0.80$. In each scenario, we further explore two different settings. In Setting (I), we fix the non-zero effects of both $\beta_{mj}$ and $\alpha_{aj}$ to be 0.5, with their signs randomly assigned as positive or negative. In Setting (II), we fix 40\% of the non-zero $\beta_{mj}$ (or $\alpha_{aj}$) to be 0.3, 30\% of them to be 0.5, and 30\% of them to be 0.7, with their signs randomly assigned as positive or negative. In both settings, we simulate the continuous exposure $\{ A_i, i = 1, ..., n \} $ independently from a standard normal distribution \textnormal{N}(0, 1). We simulate the residual error $\epsilon_{Yi}$ in the outcome model independently from \textnormal{N}(0, 1), and simulate the residual errors $\boldsymbol{\epsilon}_{Mi}$ in the mediator model from \textnormal{MVN}($\boldsymbol{0}$, ${\boldsymbol{\Sigma}}$). Here, we use the sample covariance estimated from MESA data to serve as ${\boldsymbol{\Sigma}}$ in the simulations. Afterwards, we generate a $p$-vector of mediators for the $i$th individual from $\boldsymbol{M}_i = A_i\boldsymbol{\alpha_a} + \boldsymbol{\epsilon}_{Mi}$.
We also generate the outcome $Y_i$ for the $i$th individual from $Y_i = \boldsymbol{M}_i^\top\boldsymbol{\beta_m} + A_i\beta_a + \epsilon_{Yi}$, with $\beta_a = 0.5$. 

\noindent\textit{Random Effect Simulations} In the above settings, we have fixed the effect sizes to specific values across replicates. To further examine the performance of our methods over a wide range of effect sizes, we perform additional simulations where we simulate $[
\beta_{mj}, \alpha_{aj} ]^\top$ randomly in each simulation replicate. Specifically, we generate these two effects from three different joint distributions detailed below (Figure S1): the first two correspond to the prior distributions assumed in PTG and GMM, respectively, while the last one is a horseshoe distribution, i.e.
\\
\indent (A) Simulate effects under the PTG model: $[
\beta_{mj}, \alpha_{aj} ]^\top \sim \operatorname{PTG}(\operatorname{diag}\left\{\sigma_{u}^{2}, \sigma_{u}^{2}\right\}, \lambda)$, where $\lambda=\left(\lambda_{0}, \lambda_{1}, \lambda_{2}\right)$ are set to satisfy the desired proportions of the four groups ($\pi_{1}$, $\pi_{2}$, $\pi_{3}$, $\pi_{4}$). We set $\sigma_u^2 = 0.3$ for $p = 200$, and $\sigma_u^2 = 0.1$ for $p = 2000$. \\
\indent (B) Simulate effects under the GMM model:  
  $
        \begin{bmatrix}
        \beta_{mj} \\
        \alpha_{aj} \end{bmatrix} \sim \pi_{1}\mathrm{MVN}(\boldsymbol{0}, \begin{bmatrix}
        \sigma^2 & \sigma^2/3 \\
        \sigma^2/3 & \sigma^2 \end{bmatrix}) + \pi_{2}\mathrm{MVN}(\boldsymbol{0}, \begin{bmatrix}
        \sigma^2 & 0 \\
        0 & 0 \end{bmatrix}) + \pi_{3}\mathrm{MVN}(\boldsymbol{0}, \begin{bmatrix}
        0 & 0 \\
        0 & \sigma^2 \end{bmatrix}) + \pi_{4}\boldsymbol{\delta_0}.
    $ 
    We set $\sigma^2 = 0.3$ for $p = 200$, and $\sigma^2 = 0.1$ for $p = 2000$. \\
\indent (C) Simulate effects from a mixture of bivariate horseshoe distributions, which can be generated from a scale mixture of normals:
    $
        \begin{bmatrix}
        \beta_{mj} \\
        \alpha_{aj} \end{bmatrix} \sim \pi_{1}\mathrm{MVN}(\boldsymbol{0}, 
        Z_j^2\begin{bmatrix}
        \sigma^2 & \sigma^2/3 \\
        \sigma^2/3 & \sigma^2 \end{bmatrix}) +  \pi_{2}\mathrm{MVN}(\boldsymbol{0}, Z_j^2\begin{bmatrix}
        \sigma^2 & 0 \\
        0 & 0 \end{bmatrix})
        + \pi_{3}\mathrm{MVN}(\boldsymbol{0}, Z_j^2\begin{bmatrix}
        0 & 0 \\
        0 & \sigma^2 \end{bmatrix}) + \pi_{4}\boldsymbol{\delta_0}
    $. Here, $Z_j \sim $ halfCauchy$(0,1)$, but truncated at a value of $b$ to avoid impractically large values. We set $\sigma^2 = 0.5$ for $p = 200$, and $\sigma^2 = 0.3$ for $p = 2000$, and $b=3$. Note that the effect size distribution assumed here is different from either of our proposed models, thus allowing us to study the robustness of our methods. With the effect size distributions, we follow the same procedure described as in the fixed effects settings.

We apply different methods to analyze the simulated data. In GMM, we set the Dirichlet parameters $a_{1} = 0.01p$, $a_{2} = a_{3} = 0.05p$, $a_{4} = 0.89p$. We adopt an empirical Bayesian approach to set the diagonal entries of $\boldsymbol{\Psi_0}$ as the sample variances of the non-zero $\boldsymbol{\beta_m}$ and $\boldsymbol{\alpha_a}$ fitted through Lasso. We set the degree of freedom $\nu$ in the inverse-Wishart distribution to be two, which makes the distribution reasonably non-informative while still well-defined. In PTG, we set the pre-defined minimal detectable effect sizes ($\lambda_0, \lambda_1, \lambda_2$) to be the 90\% quantiles of the estimated $|\boldsymbol{\beta_m}|$ and $|\boldsymbol{\alpha_a}|$ fitted through Lasso. To be consistent with the GMM, we choose the parameter $\hat{\tau}^2$ in the priors $\tau_{\beta}^2 \sim \mathrm{IG}(1.1, \hat{\tau}^2), \tau_{\alpha}^2 \sim \mathrm{IG}(1.1, \hat{\tau}^2)$ to ensure that the prior inclusion probability is around 0.01. 
For the Bayesian methods, we perform 150,000 iterations and discard the first 100,000 iterations as burn-in. We check the MCMC convergence by running five chains with random initial values and calculating the potential scale reduction factor (PSRF) for the PIPs. All the PSRFs fall within (1.0, 1.2), indicating the convergence of our algorithms. 

\noindent\textit{\textbf{Results for Fixed Effect Simulations: Setting (I)-(II)}} Table \ref{tbl:power0102} shows the results under the fixed effects for the small sample scenario $n = 100, p = 200$ and the large sample scenario $n = 1000, p = 2000$. Overall, our proposed methods, GMM and PTG, outperform the other methods. These two methods achieve the highest AUC and are up to $\sim$ 30\% more powerful than the other methods in identifying active mediators, with performance gain more apparent in the large sample scenario. Under Setting (I) where the mediation effects are large, the PTG method has the highest average TPR for both small and large sample settings. The performance of PTG is followed by GMM and BAMA. In contrast, under Setting (II) where the mediation effects are uneven, PTG may fail to identify some of the active mediators with small effects due to the thresholding set by the pre-defined parameter $\lambda$. Instead, GMM performs the best and its performance is followed by PTG and BAMA. Importantly, median inclusion probabilities (PIP = 0.5) in both GMM and PTG can be used as a criterion to declare active mediators (details in SM), producing decent empirical estimates for FDR in simulations (Table S1, S2). Among the frequentist methods, the Bi-Lasso performs best over the others and is also competitive in the small sample setting. HIMA and the univariate method are among the worst methods for mediator selection, presumably because neither models the entire set of mediators jointly in the outcome model. 

In terms of the effects estimation, GMM has the lowest MSE$_{\textnormal{non-null}}$ across most simulation scenarios. Due to hard thresholding, PTG tends to provide a conservative list of the active mediators. Consequently, the non-zero indirect effects of some active mediators are shrunk to zero in PTG, leading to relatively high MSE$_{\textnormal{non-null}}$ but small MSE$_{\textnormal{null}}$ by PTG. Meanwhile, we find that the Pathway Lasso does not appear to exhibit much advantage over the simple alternative Bi-Lasso. Indeed, Pathway Lasso requires multiple tuning parameters for inducing the penalty term on the indirect effects, and those tuning parameters may benefit from more careful specifications than the algorithm default setting. The univariate method in particular has a quite high MSE$_{\textnormal{null}}$ as it does not apply any shrinkage on the effect estimates. Overall, by jointly analyzing multiple mediators in a coherent statistical framework, both PTG and GMM outperform the other methods in simulations. 

\begin{table}
\caption{\label{tbl:power0102} Simulation results for fixed effects under $n=100, p = 200$ and $n=1000, p=2000$, $p_{11}$ is the number of truly active mediators. TPR:  true positive rate at false discovery rate (FDR) = 0.10. MSE$_{\textnormal{non-null}}$: mean squared error for the indirect effects of truly active mediators. MSE$_{\textnormal{null}}$: mean squared error for the indirect effects of truly inactive mediators. The results are based on 200 replicates for each setting, and the standard errors are shown within parentheses. For PTG, we include the pre-defined thresholds $(\lambda_0, \lambda_1, \lambda_2)$ under each setting. Bolded TPRs indicate the top two performers.}
\centering{
  \begin{tabular}{*{5}{c}}
    \hline    
    \multicolumn{5}{c}{$n = 100, p = 200, p_{11} = 10$, \textit{fixed effects (I)}} \\
    \hline
    \em Method & \em AUC & \em TPR & \em MSE$_{\textnormal{non-null}}$ & \em MSE$_{\textnormal{null}} \times 10^{-4}$ \\
    \hline
    PTG (0.15,0.4,0.4) & 0.99(0.001) & $\boldsymbol{0.54}$(0.025) & 0.040 & 0.486  \\
    GMM & 0.98(0.002) & $\boldsymbol{0.42}$(0.023) & 0.054 & 1.336 \\
    BAMA & 0.98(0.001) & 0.40(0.022) & 0.054 & 1.870  \\
    Bi-BLasso & 0.90(0.005) & 0.27(0.015) & 0.092 & 21.879  \\
    PathLasso & 0.75(0.004) & 0.35(0.023) & 0.098 & 17.220 \\
    Bi-Lasso & 0.81(0.008) & 0.38(0.019) & 0.079 & 12.436  \\
    HIMA & 0.61(0.005) & 0.23(0.010) & 0.081 & 1.832  \\
    Univariate & 0.83(0.007) & 0.25(0.014) & 0.088 & 26.220  \\
	\hline
	\multicolumn{5}{c}{$n = 100, p = 200, p_{11} = 10$, \textit{fixed effects (II)}} \\
    \hline
    \em Method & \em AUC & \em TPR & \em MSE$_{\textnormal{non-null}}$ & \em MSE$_{\textnormal{null}} \times 10^{-4}$ \\
	\hline
    PTG (0.15,0.4,0.4) & 0.96(0.003) & $\boldsymbol{0.34}$(0.017) & 0.074 & 0.549  \\
    GMM & 0.96(0.003) & $\boldsymbol{0.39}$(0.020) & 0.029 & 0.791 \\
    BAMA & 0.96(0.003) & 0.31(0.015) & 0.038 & 1.502 \\
    Bi-BLasso & 0.90(0.005) & 0.25(0.013) & 0.044 & 11.040 \\
    PathLasso & 0.72(0.007) & 0.23(0.011) & 0.072 & 3.225 \\
    Bi-Lasso & 0.72(0.006) & 0.30(0.014) & 0.041 & 0.445  \\
    HIMA & 0.56(0.005) & 0.17(0.009) & 0.056 & 2.526  \\
    Univariate & 0.81(0.006) & 0.19(0.013) & 0.030 & 46.764 \\
    \hline
    \multicolumn{5}{c}{$n = 1000, p = 2000, p_{11} = 100$, \textit{fixed effects (I)}} \\
    \hline
   \em Method & \em AUC & \em TPR & \em MSE$_{\textnormal{non-null}}$ & \em MSE$_{\textnormal{null}} \times 10^{-4}$ \\
    \hline
    PTG (0.15,0.4,0.4)  & 0.98(0.001) & $\boldsymbol{0.64}$(0.008) & 0.028 & 0.070 \\
    GMM & 0.99(0.001) & $\boldsymbol{0.61}$(0.009) & 0.023 & 0.170  \\
    BAMA & 0.98(0.001) & 0.57(0.010) & 0.038 & 0.141 \\
    Bi-BLasso & 0.90(0.002) & 0.23(0.004) & 0.063 & 5.711 \\
    PathLasso & 0.70(0.002) & 0.20(0.005) & 0.057 & 3.982 \\
    Bi-Lasso & 0.76(0.001) & 0.25(0.003) & 0.051 & 0.290 \\
    HIMA & 0.57(0.001) & 0.16(0.003) & 0.077 & 1.891 \\
    Univariate & 0.93(0.001) & 0.10(0.005) & 0.092 & 225.056  \\
	\hline
	\multicolumn{5}{c}{$n = 1000, p = 2000, p_{11} = 100$, \textit{fixed effects (II)}} \\
	\hline
	\em Method & \em AUC & \em TPR & \em MSE$_{\textnormal{non-null}}$ & \em MSE$_{\textnormal{null}} \times 10^{-6}$ \\
    \hline
    PTG (0.15,0.4,0.4) & 0.90(0.002) & $\boldsymbol{0.40}$(0.008) & 0.008 & 0.164 \\
    GMM & 0.97(0.001) & $\boldsymbol{0.48}$(0.006) & 0.003 & 3.257 \\
    BAMA & 0.96(0.001) & 0.36(0.007) & 0.005 & 7.346 \\
    Bi-BLasso & 0.85(0.001) & 0.18(0.004) & 0.011 & 184.761 \\
    PathLasso & 0.67(0.002) & 0.19(0.003) & 0.017 & 19.540 \\
    Bi-Lasso & 0.70(0.001) & 0.23(0.005) & 0.007 & 4.925 \\
    HIMA & 0.56(0.002) & 0.09(0.004) & 0.013 & 23.048 \\
    Univariate & 0.90(0.002) & 0.12(0.003) & 0.075 & 208.660 \\
    \hline
\end{tabular}}
\end{table}

\noindent\textit{\textbf{Results for Random Effect Simulations: Setting (A)-(C)}} Table \ref{tbl:power1} shows the results in the small sample scenario and Table \ref{tbl:power2} shows the results in the large sample scenario. In all the settings, our proposed methods, PTG and GMM, outperform the other methods with an approximately 10\% power gain in identifying active mediators. Between PTG and GMM, we find that both methods work preferably well in the setting where their corresponding effect size distribution is used. Specifically, in Setting (A) with $p = 2000$, the PTG method has the highest AUC (0.98) and TPR (0.40) at FDR = 10\%. The performance of PTG is followed by GMM (AUC = 0.98, TPR = 0.37). In Setting (B) with $p = 2000$, the GMM method has the highest AUC (0.95) and TPR (0.51). The performance of GMM is followed by PTG (AUC = 0.92, TPR = 0.42). In Setting (C) where the effects are simulated with a horseshoe distribution, we find that GMM performs the best and its performance is followed by PTG and BAMA. The horseshoe distribution has a tall spike near zero and heavy tails, and therefore leads to a particularly challenging setting for most methods. The good performance of GMM in Setting (C) thus supports the robustness of the method. In addition, as before, both PTG and GMM provide reasonable empirical estimates of FDR and TPR (Table S1, S2 of SM) based on a PIP = 0.5 cutoff. The accuracy gain in indirect effects estimation basically follows the same pattern as the power gain in mediator selection. The computing time of the proposed methods is reported in Table S3 of SM, with both methods being relatively efficient for $p=200$ and $p=2000$ cases. 

Finally, among the three frequentist methods, the bi-Lasso yields higher power as compared to the other two in all the scenarios and has smaller MSE in almost all the settings except for the horseshoe setting. 
Between bi-Lasso and bi-Bayesian Lasso, we find that the former outperforms the latter with higher TPR and smaller MSE$_{\textnormal{null}}$. This comparison between bi-Lasso and bi-Bayesian Lasso suggests that under this sparse setup, the estimated non-sparse indirect effects in bi-Bayesian Lasso may not be ideal for classifying mediators as compared to the sparse solutions produced by bi-Lasso. 

In summary, the simulations demonstrate that GMM enjoys superior and robust performance for mediator selection and effect estimation, while PTG is preferable under potentially large non-zero effects in mediator selection.

\begin{table}
\caption{Simulation results for random effects under $n=100, p = 200$, $p_{11}$ is the number of truly active mediators. TPR:  true positive rate at false discovery rate (FDR) = 0.10. MSE$_{\textnormal{non-null}}$: mean squared error for the indirect effects of truly active mediators. MSE$_{\textnormal{null}}$: mean squared error for the indirect effects of truly inactive mediators. The results are based on 200 replicates for each setting, and the standard errors are shown within parentheses. For PTG, we include the pre-defined thresholds $(\lambda_0, \lambda_1, \lambda_2)$ under each setting. Bolded TPRs indicate the top two performers.\label{tbl:power1}}
\centering
  \begin{tabular}{*{5}{c}}
    \hline      
    \multicolumn{5}{c}{$n = 100, p = 200, p_{11} = 10, PTG, \sigma_{u}^2 = 0.3$} \\
    \hline
    \em Method & \em AUC & \em TPR & \em MSE$_{\textnormal{non-null}}$ & \em MSE$_{\textnormal{null}} \times 10^{-4}$ \\
    \hline
    PTG (0.15, 0.4, 0.4) & 0.98(0.002) & $\boldsymbol{0.45}$(0.020) & 0.05 & 1.59  \\
    GMM & 0.98(0.001) & $\boldsymbol{0.43}$(0.015) & 0.03 & 4.25 \\
    BAMA & 0.98(0.001) & 0.41(0.019) & 0.04 & 2.64  \\
    Bi-BLasso & 0.89(0.006) & 0.35(0.017) & 0.05 & 6.83 \\
    PathLasso & 0.65(0.013) & 0.31(0.015) & 0.06 & 2.43 \\
    Bi-Lasso & 0.78(0.009) & 0.40(0.020) & 0.05 & 1.12 \\
    HIMA & 0.60(0.007) & 0.29(0.012) & 0.07 & 5.46 \\
    Univariate & 0.85(0.008) & 0.29(0.023) & 0.15 & 76.25 \\
    \hline
	 \multicolumn{5}{c}{$n = 100, p = 200, p_{11} = 10, Gaussian, \sigma^2 = 0.3$} \\
	 \hline
	 \em Method & \em AUC & \em TPR & \em MSE$_{\textnormal{non-null}} \times 10^{-3}$ & \em MSE$_{\textnormal{null}} \times 10^{-5}$ \\
    \hline
    PTG (0.04, 0.2, 0.2) & 0.92(0.002) & $\boldsymbol{0.38}$(0.008) & 6.24 & 4.05 \\
    GMM & 0.94(0.003) & $\boldsymbol{0.41}$(0.006) & 3.92 & 3.56 \\
    BAMA & 0.95(0.003) & $\boldsymbol{0.38}$(0.011) & 5.06 & 3.39  \\
    Bi-BLasso & 0.83(0.006) & 0.28(0.014) & 23.31 & 14.38 \\
    PathLasso & 0.75(0.008) & 0.30(0.011) & 11.57 & 3.09 \\
    Bi-Lasso & 0.75(0.003) & 0.36(0.011) & 7.50 & 1.52 \\
    HIMA & 0.65(0.005) & 0.21(0.009) & 14.98 & 7.93 \\
    Univariate & 0.75(0.006) & 0.26(0.025) & 62.46 & 234.30 \\
    \hline
    \multicolumn{5}{c}{$n = 100, p = 200, p_{11} = 10, Horseshoe, \sigma^2 = 0.5, b = 3$} \\
    \hline
    \em Method & \em AUC & \em TPR & \em MSE$_{\textnormal{non-null}}$ & \em MSE$_{\textnormal{null}} \times 10^{-4}$ \\
    \hline
    PTG (0.15, 0.5, 0.3) & 0.80(0.009) & $\boldsymbol{0.30}$(0.015) & 0.42 & 7.16 \\
    GMM & 0.83(0.006) & $\boldsymbol{0.33}$(0.011) & 0.03 & 5.21 \\
    BAMA & 0.80(0.008) & 0.28(0.017) & 0.11 & 6.28 \\
    Bi-BLasso & 0.76(0.011) & 0.23(0.010) & 0.45 & 42.36 \\
    PathLasso & 0.65(0.019)  & 0.25(0.026) & 0.51 & 6.04 \\
    Bi-Lasso & 0.68(0.009) & 0.27(0.017) & 0.46 & 5.41 \\
    HIMA & 0.60(0.006) & 0.20(0.010) & 0.41 & 26.51 \\
    Univariate & 0.72(0.009) & 0.20(0.020) & 0.44 & 512.33 \\
    \hline
\end{tabular}
\end{table}

\begin{table}
\caption{\label{tbl:power2} Simulation results for random effects under $n = 1000, p = 2000$, $p_{11}$ is the number of truly active mediators. TPR:  true positive rate at false discovery rate (FDR) = 0.10, MSE$_{\textnormal{non-null}}$: mean squared error for the indirect effects of truly active mediators. MSE$_{\textnormal{null}}$: mean squared error for the indirect effects of truly inactive mediators. The results are based on 200 replicates for each setting, and the standard errors are shown within parentheses. For PTG, we include the pre-defined thresholds $(\lambda_0, \lambda_1, \lambda_2)$ under each setting. Bolded TPRs indicate the top two performers.}
\centering
  \begin{tabular}{*{5}{c}}
    \hline  
 	\multicolumn{5}{c}{$n = 1000, p = 2000, p_{11} = 100, PTG, \sigma_u^2 = 0.1$ } \\
 	\hline
    \em Method & \em AUC & \em TPR & \em MSE$_{\textnormal{non-null}} \times 10^{-4}$ & \em MSE$_{\textnormal{null}} \times 10^{-6}$ \\
    \hline
    PTG (0.05,0.15,0.15) & 0.98(0.001) & $\boldsymbol{0.40}$(0.008) & 5.28 & 2.46 \\
    GMM & 0.98(0.001) & $\boldsymbol{0.37}$(0.010) & 3.86 & 4.26 \\
    BAMA & 0.98(0.001) & 0.30(0.012) & 4.84 & 3.62 \\
    Bi-BLasso & 0.92(0.003) & 0.29(0.018) & 7.92 & 11.38 \\
    PathLasso & 0.77(0.009) & 0.22(0.007) & 7.02 & 1.74 \\
    Bi-Lasso & 0.83(0.003) & 0.28(0.014) & 5.60 & 1.81 \\
    HIMA & 0.53(0.002) & 0.14(0.004) & 9.96 & 4.96 \\
    Univariate & 0.85(0.003) & 0.11(0.023) & 60.24 & 214.57 \\
    \hline
    \multicolumn{5}{c}{$n = 1000, p = 2000, p_{11} = 100, Gaussian, \sigma^2 = 0.1$} \\
    \hline
    \em Method & \em AUC & \em TPR & \em MSE$_{\textnormal{non-null}} \times 10^{-3}$ & \em MSE$_{\textnormal{null}} \times 10^{-5}$  \\
    \hline
    PTG (0.02,0.2,0.1) & 0.92(0.002) & $\boldsymbol{0.42}$(0.006) & 4.76 & 0.874 \\
    GMM & 0.95(0.001) & $\boldsymbol{0.51}$(0.007) & 2.09 & 0.712 \\
    BAMA & 0.90(0.003) & 0.41(0.018) & 2.85 & 0.722 \\
    Bi-BLasso & 0.88(0.002) & 0.32(0.007) & 4.85 & 1.632 \\
    PathLasso & 0.78(0.011) & 0.25(0.003) & 4.88 & 1.256 \\
    Bi-Lasso & 0.81(0.002) & 0.38(0.010) & 2.53 & 0.368 \\
    HIMA & 0.55(0.002) & 0.19(0.004) & 8.41 & 1.544 \\
    Univariate & 0.82(0.003) & 0.19(0.017) & 34.08 & 20.05 \\
    \hline
    \multicolumn{5}{c}{$n = 1000, p = 2000, p_{11} = 100, Horseshoe, \sigma^2 = 0.3, b = 3$} \\
    \hline
    \em Method & \em AUC & \em TPR & \em MSE$_{\textnormal{non-null}}$ & \em MSE$_{\textnormal{null}} \times 10^{-4}$  \\
    \hline
    PTG (0.03,0.3,0.1) & 0.74(0.002) & $\boldsymbol{0.29}$(0.008) & 0.18 & 10.04 \\
    GMM & 0.80(0.001) & $\boldsymbol{0.38}$(0.007) & 0.14 & 2.94 \\
    BAMA & 0.75(0.002) & 0.27(0.006) & 0.25 & 3.88 \\
    Bi-BLasso & 0.71(0.002) & 0.09(0.003) & 0.26 & 127.55 \\
    PathLasso & 0.66(0.008) & 0.05(0.002) & 0.41 & 2.03 \\
    Bi-Lasso & 0.72(0.003) & 0.24(0.007) & 0.24 & 1.57 \\
    HIMA & 0.55(0.002) & 0.09(0.004) & 0.39 & 1.56 \\
    Univariate & 0.77(0.003) & 0.09(0.015) & 0.59 & 644.07 \\
    \hline
\end{tabular}
\end{table}

\section{Data Application}
\label{sec6}

\subsection{Analysis of DNA Methylation in the MESA Cohort}
We applied the proposed GMM and PTG to investigate the mediation mechanism of DNAm in the pathway from neighborhood socioeconomic disadvantage to BMI in the MESA data. Neighborhood SES is the exposure variable and is created based on a principal components analysis of 16 census-tract level variables reflecting dimensions of education, occupation, income, poverty, housing, etc. 
BMI is the outcome variable and also a critical risk factor for various diseases including T2D and CVD (\citealp{hjellvik2012body}). Understanding how methylation at different CpG sites mediates the effects of neighborhood SES on BMI can shed light on the molecular mechanisms of complex diseases, thus leading to potential therapeutic strategies. The detailed processing steps for MESA data are provided in the SM. Briefly, we selected 1,225 individuals with non-missing data. Due to computational reasons, we focused on a final set of 2,000 CpG sites that have the strongest marginal associations with neighborhood SES. We applied various methods for the mediation analysis. In the outcome model, we adjust for age, gender, race/ethnicity, childhood socioeconomic status (SES) and adult SES. In the mediator model, we control for age, gender, race/ethnicity, childhood SES, adult SES, and enrichment scores for 4 major blood cell types (neutrophils, B cells, T cells and natural killer cells). All the continuous variables are standardized to have zero mean and unit variance.

\begin{figure}
\centering \makebox{\includegraphics[scale=0.33]{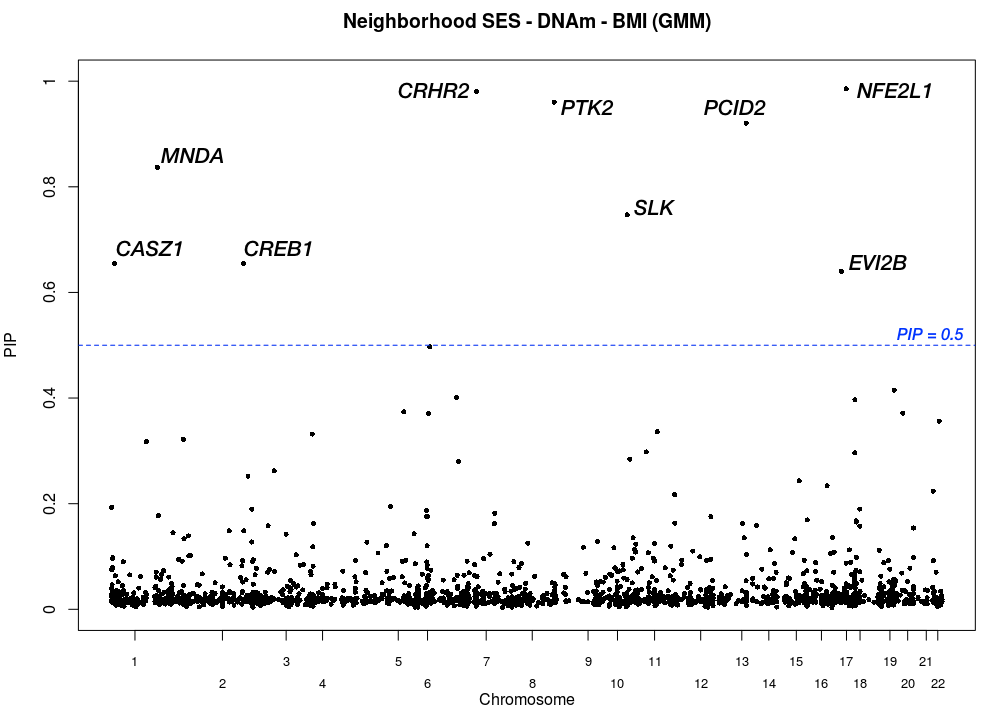}}
\makebox{\includegraphics[scale=0.33]{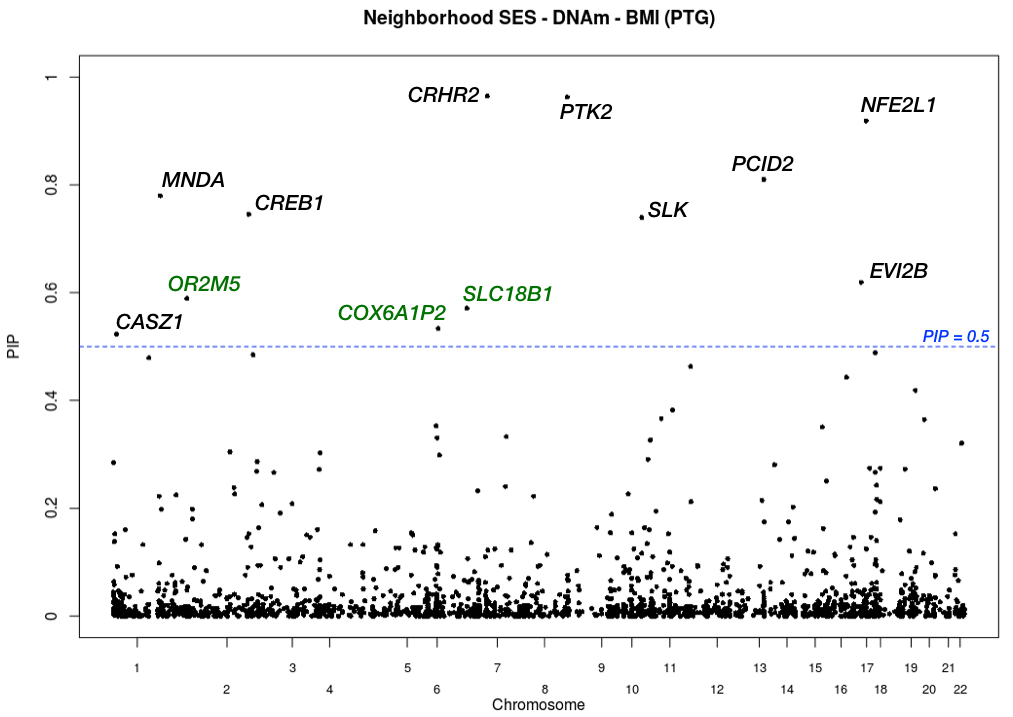}}
\caption{Data analysis results for the trio \textcolor{black}{Neighborhood SES} $\rightarrow$ DNAm $\rightarrow$ \textcolor{black}{BMI} in MESA data. The upper panel shows the PIPs obtained from the GMM method, and the lower panels shows the PIPs obtained from the PTG method. The blue lines mark the PIP = 0.5 threshold, and we include the nearby genes of the selected CpG sites. Most of the sites are identified by both methods, and the three genes in green are additional findings from PTG.}
\label{fig:pip1}
\end{figure}

We display the PIP values for each of the 2,000 CpG sites from PTG and GMM in Figure \ref{fig:pip1}. GMM identified nine CpG sites with significant evidence for mediating the neighborhood SES effects on BMI based on 0.5 cutoff of PIPs. In contrast, PTG identified twelve significant CpG sites at the same threshold, which include all the nine sites selected by GMM method. The top five CpG sites identified by the two methods are identical. The rank correlation for the mediator rank lists obtained from both methods is 0.87, supporting the high consistency between the two methods. We carefully examine the nearby genes of the detected methylation sites by GMM and PTG. Among them, the protein-coding gene \textit{PTK2} has been previously discovered as BMI risk loci (\citealp{zeller2018impact}); \textit{PCID2} and \textit{NFE2L1} have been shown to be associated with obesity, glucose, diabetes and related metabolic diseases (\citealp{zheng2015cnc, erdmann2018decade}); \textit{COX6A1P2} was robustly recognized to link with obesity development in multiple epigenome-wide studies (\citealp{kvaloy2018epigenome}) and \textit{EVI2B} was reported as one of the regulatory genes related to obesity (\citealp{kogelman2014identification}). Therefore, the genes nearby the detected CpG sites may play an important role in transmitting the effects of neighborhood SES to BMI. For the other competing methods, BAMA and the univariate methods do not have sufficient power to identify any significant CpG sites at 0.10 FDR. HIMA identifies one CpG site in the gene region of \textit{PCID2} as active mediator through its joint significance test (adjusted $P$-value = 6.3e-5), and this single site has also been detected by PTG and GMM methods. Bi-Lasso and Pathway Lasso tend to produce a large number of false positives in simulations, and thus it is hard to verify their findings in the real data application.

\subsection{Analysis of Endogenous Biomarkers and Environmental Data in the LIFECODES Birth Cohort} 
As another data example, we study the collective impact of endogenous signaling molecules derived from lipids, peptides, and DNA in mediating prenatal exposure to environmental contaminants on the risk of preterm birth in the LIFECODES birth cohort. Detailed description of the study is provided in the SM. Briefly, we consider $n = 161$ pregnant women registered at the Brigham and Women's Hospital in Boston, MA between 2006 and 2008. Subjects' urine and plasma specimens were collected at one study visit occurring between 23.1 and 28.9 weeks gestation. Four classes of environmental contaminants, including phthalates, phenols, polycyclic aromatic hydrocarbons, and trace metals, were measured in each urine sample. Among them, phthalates are the high-production volume chemicals commonly used as plasticizers in numerous consumer products. Previous studies have shown that everyday exposure to phthalates during pregnancy would increase risk of delivering preterm (\citealp{ferguson2014variability}). Recent studies have also uncovered associations between multiple lipid biomarkers and preterm birth (\citealp{aung2019prediction}). Based on these previous literature, we aim to understand the molecular mechanism underlying the effects of phthalates on preterm. To do so, we follow \citet{Aung2020.05.30.20117655} to create an environmental risk score for the phthalate class and treat such risk score as the exposure variable. We recorded the gestational age at delivery as the continuous birth outcome. In terms of mediators, we obtained 61 endogenous biomarkers from urine and plasma that included 51 eicosanoids, five oxidative stress biomarkers and five immunological biomarkers. With these variables, we examine if any of these 61 available biomarkers mediates the effects of phthalate exposure on gestational age at delivery. In the analysis, we perform log-transformation on all measurements of the exposure metabolites and endogenous biomarkers. We adjust for age and maternal BMI from the initial visit, race, and urinary specific gravity levels inside both models of the mediation analysis. Since the cohort is oversampled for preterm cases ($<$ 37 weeks gestation), we multiply the data by the case-control sampling weights to adjust for that. 

\begin{figure}[!h]
\centering 
\makebox{\includegraphics[scale=0.42]{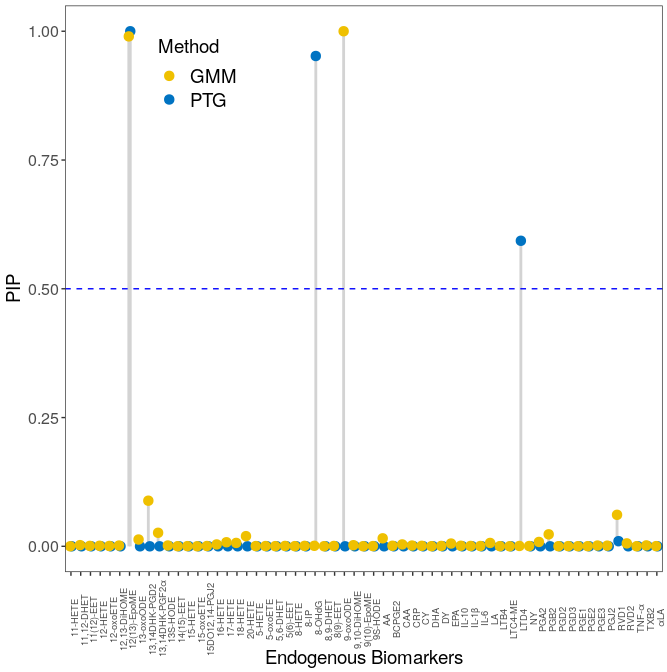}}
\caption{Data analysis results for the LIFECODES cohort. The panel shows the PIPs obtained from GMM (blue) and PTG (yellow) methods for the trio Exposure to phthalates $\rightarrow$ Biomarkers $\rightarrow$ Gestational age of the newborn at delivery. }
\label{fig:pip2}
\end{figure}

\parskip 0.1in
We applied the proposed methods to the data and summarize the results in Table \ref{tbl:res2}. Both PTG and GMM identified significant mediators that mediate the effects of the phthalate exposure on gestational age at delivery based on PIP = 0.5 cutoff (Figure \ref{fig:pip2}), with rank lists of mediators positively correlated with each other (rank correlation = 0.48). Specifically, GMM identified two significant biomarkers (9-oxooctadeca-dienoic acid [9-oxoODE], 12,13-epoxy-octadecenoic acid
[12(13)-EpoME]). PTG identified three significant biomarkers (8-hydroxydeoxyguanosine
[8-OHdG], 12(13)-EpoME, leukotriene D4 [LTD4]), one of which (12(13)-EpoME) overlaps with those identified by GMM. Among the identified biomarkers, 8-OHdG is commonly utilized as a marker of oxidative stress generated upon repair of oxidative DNA damage and has been found to be strongly associated with decreased gestational length and increased risk of preterm (\citealp{ferguson2015repeated}); while LTD4 has been shown to exhibit significant associations with preterm birth, and 9-oxoODE and 12(13)-EpoME had an important protective effect on preterm birth (\citealp{aung2019prediction}). As a comparison, BAMA, HIMA and the univariate methods fail to identify any significant active mediators at 0.10 FDR in this application. Our results help improve the understanding of the molecular mechanisms underlying the effects of environmental exposure on preterm, and could further lead to improvement of treatment and prevention strategies. 

\begin{table}
\caption{\label{tbl:res2}Summary of the identified active mediators from the data application on MESA and LIFECODES study. For PTG, we include the pre-defined thresholds $(\lambda_0, \lambda_1, \lambda_2)$ for the two real datasets.}
\centering
  \begin{tabular}{*{2}{c}}
  \hline
	\em Method & \em Selected Mediators \\
	\hline
    \multicolumn{2}{c}{\textit{MESA: Neighborhood SES $\rightarrow$ DNAm $\rightarrow$ BMI}} \\
    \hline
    GMM  & CRHR2, NFE2L1, PTK2, PCID2, MNDA, \\
         & SLK, CREB1, CASZ1, EVI2B \\
    PTG (0.01,0.05,0.1) & CRHR2, NFE2L1, PTK2, PCID2, MNDA, CREB1, \\
        & SLK, EVI2B, OR2M5, SLC18B1, COX6A1P2, CASZ1 \\
    \hline
    \multicolumn{2}{c}{\textit{LIFECODES: Phthalates $\rightarrow$ Biomarkers $\rightarrow$ \textit{Gestational age}}} \\
    \hline
    GMM  & 12(13)-EpoME, 9-oxoODE \\
    PTG (3.0,2.0,1.5) & 12(13)-EpoME, 8-OHdG, LTD4 \\
    \hline
\end{tabular}
\end{table}

\section{Discussion}
\label{sec7}

In this paper, we present two novel joint modeling approaches, PTG and GMM, for high-dimensional mediation analysis. Our methods can jointly model a large number of mediators and enable penalization on the indirect effects in a targeted way. Our methods effectively characterize the high-dimensional set of potential mediators into four groups based on the exposure-mediator and mediator-outcome effects: the active mediating group and three non-mediating groups. These group categorizations are in consonance with the composite structure for testing the indirect effect recently proposed in genome-wide mediation analyses (\citealp{huang2019genome}). With extensive simulations, we show that our methods achieve up to 30\% power gain in identifying true non-null mediators as compared with other existing alternatives, including several recently developed penalized and Bayesian methods for mediation analysis. We have demonstrated the benefits of our methods with both genetic and environmental data in the MESA and LIFECODES cohorts. For example, in the MESA cohort, we identify several DNAm and their nearby genes, e.g. \textit{NFE2L1} and \textit{PTK2}, with strong evidence for mediating neighborhood SES effects on BMI. This is an important finding in biosocial research where we try to characterize how the insults from our external environment influence the internal cellular environment and finally manifest into development of a chronic disease.

On the methodological front, we still have challenges remaining that are unsolved in this modeling exercise. Bayesian FDR control is of great importance to safeguard false positives in the scientific discovery. For PTG and GMM, we rely on the median inclusion probabilities (PIP = 0.5) to identify active mediators, which provides effective FDR control as validated through simulations. For bi-Bayesian Lasso and other continuous shrinkage methods, such as the scale mixture of normals prior, we have attempted to define PIP using shrinkage factors following \citet{carvalho2010horseshoe}. However, we find it challenging to adapt the shrinkage factors to devise an optimal strategy for computing PIP analogs and ranking correlated mediators. Consequently, we have to rely on the estimated indirect effects from these methods to rank mediators, which may account for at least partially the relatively poor performance of these methods. Therefore, coming up with an analog of PIP as the selection criterion 
in mediation analysis for various other methods remains a topic of future investigation. 


One limitation of our current work is that the proposed methods do not explicitly incorporate the correlation structure among mediators in the modeling process. Treating mediators independent \textit{a priori}, the models may fail to distinguish among highly correlated mediators and lose power in mediator selection when two truly active mediators tend to be correlated with each other. Correlations among mediators are commonly seen in modern data analysis; such examples include genomic data that measure hundreds of thousands of gene expressions/single nucleotide polymorphisms (SNPs), and brain image data that contain a large number of voxels/regions. Incorporating mediator correlation information explicitly into our Bayesian paradigm could be a promising direction for future work.

\section{Software}
Software in the form of C++ codes is available on github \url{https://github.com/yanys7/GMM_PTG_Mediation}. 

\section*{Acknowledgments}

This work was supported by NSF DMS1712933 (B.M., X.Z.), NIH R01HG009124 (X.Z.), NIH R01HL141292 (J.S.), and NIH R01MD011724 (B.N.). MESA and the MESA SHARe project are conducted and supported by the National Heart, Lung, and Blood Institute (NHLBI) in collaboration with MESA investigators. Support for MESA is provided by contracts HHSN268201500003I, N01-HC-95159, N01-HC-95160, N01-HC-95161, N01-HC-95162, N01-HC-95163, N01-HC-95164, N01-HC-95165, N01-HC-95166, N01-HC-95167, N01-HC-95168, N01-HC-95169, UL1-TR-000040, UL1-TR-001079, UL1-TR-001420, UL1-TR-001881, and DK063491. The MESA Epigenomics \& Transcriptomics Study was funded by NHLBI, NIA, and NIDDK grants: 1R01HL101250, R01 AG054474, and R01 DK101921. The authors thank the other investigators, the staff, and the participants of the MESA study for their valuable contributions. A full list of participating MESA investigators and institutions can be found at http://www.mesa-nhlbi.org.

\bibliographystyle{rss}
\bibliography{reference}

\end{document}


\date{}
\maketitle

\section{Identifiability Assumptions for Causal Mediation Analysis }
\hfill \break
We use the same counterfactual notation as in the main manuscript. To connect potential variables to observed data, we make the Stable Unit Treatment Value Assumption (SUTVA) \citep{rubin1980randomization, rubin1986comment}. Specifically, the SUTVA assumes there is no interference between subjects and the consistency assumption, which states that the observed variables are the same as the potential variables corresponding to the actually observed treatment level, i.e., $\boldsymbol{M_i}=\sum_a { \boldsymbol{M_i}(a) I(A_i=a)}$, and $Y_i=\sum_a \sum_{\boldsymbol{m}} {Y_i (a, \boldsymbol{m}) I(A_i = a, \boldsymbol{M_i} = \boldsymbol{m})}$, where $I(\cdot)$ is the indicator function. 

Causal effects are formally defined in terms of potential variables which are not necessarily observed, but the identification of causal effects must be based on observed data. Therefore further assumptions regarding the confounders are required for the identification of causal effects in mediation analysis \citep{vanderweele2014mediation}. We will use $A \indep B | C$ to denote that $A$ is independent of $B$
conditional on $C$. To estimate the average NDE and NIE from observed data, the following assumptions are needed: (1) $Y_i(a,\boldsymbol{m}) \indep A_i | \boldsymbol{C_i}$, no unmeasured confounding for exposure-outcome relationship; (2) $Y_i(a,\boldsymbol{m}) \indep \boldsymbol{M}_i | \{\boldsymbol{C_i}, A_i\}$, no unmeasured confounding for any of mediator-outcome relationship after controlling for the exposure; (3) $\boldsymbol{M}_i(a) \indep A_i | \boldsymbol{C_i}$, no unmeasured confounding for the exposure effect on all the mediators; (4) $Y_i(a,\boldsymbol{m}) \indep \boldsymbol{M}_i(a^{\star}) | \boldsymbol{C_i}$, no downstream effect of the exposure that confounds any mediator-outcome relationship. The four assumptions are required to hold with respect to the whole set of mediators. Finally, as in all mediation analysis, the temporal ordering assumption also needs to be satisfied, i.e., the exposure precedes the mediators, and the mediators precede the outcome.

\section{Posterior Sampling Algorithm Details for Gaussian Mixture Model (GMM)}

Let $\boldsymbol{\Theta}_{GMM} = (\boldsymbol{\beta_m}, \boldsymbol{\alpha_a}, \boldsymbol{V_k}, \beta_a, \boldsymbol{\beta_c}, \boldsymbol{\alpha_c}, \{ \boldsymbol{\gamma}_j\}_{j=1}^p, \pi_k, k = 1,2,3,4, \sigma_e^2, \boldsymbol{\Sigma}, \sigma_a^2)$ denote all the unknown parameters in our Gaussian mixture model. The joint likelihood of $\{ Y_i, \boldsymbol{M_i} \}_{i=1}^n$ given $\boldsymbol{\Theta}_{GMM}$ is,
\vspace{-0.5in}
\begin{adjustwidth}{-0.5cm}{}
\begin{eqnarray*}
\textnormal{log}P(\{ Y_i, \boldsymbol{M_i} \}_{i=1}^n | \boldsymbol{\Theta}_{GMM}, \{ A_i, \boldsymbol{C_i} \}_{i=1}^n) &=& \sum_{i=1}^n \textnormal{log}P(Y_i, \boldsymbol{M_i} | \boldsymbol{\Theta}_{GMM}, A_i, \boldsymbol{C_i}) \nonumber \\
&=& \sum_{i=1}^n \textnormal{log}P(Y_i | \boldsymbol{M_i}, \boldsymbol{\beta_m}, \sigma_e^2, \beta_a, \boldsymbol{\beta_c}, A_i, \boldsymbol{C_i}) \nonumber \\
&& + \textnormal{log}P(\boldsymbol{M_i} | \boldsymbol{\alpha_a}, \boldsymbol{\alpha_c}, \boldsymbol{\Sigma}, A_i, \boldsymbol{C_i}) \\
&=& \sum_{i=1}^n -\frac{1}{2}\textnormal{log}\sigma_e^2 - \frac{1}{2\sigma_e^2}(Y_i - \boldsymbol{M_i}^T\boldsymbol{\beta_m} - A_i\beta_a - \boldsymbol{C_i}^T\boldsymbol{\beta_c})^2 \nonumber \\
&-& \frac{1}{2}\textnormal{log}|\boldsymbol{\Sigma}| - \frac{1}{2}(\boldsymbol{M_i} -  A_i\boldsymbol{\alpha_a} - \boldsymbol{\alpha_c}\boldsymbol{C_i})^T \boldsymbol{\Sigma}^{-1}(\boldsymbol{M_i} -  A_i\boldsymbol{\alpha_a} - \boldsymbol{\alpha_c}\boldsymbol{C_i})
\end{eqnarray*}
\end{adjustwidth}

The joint log posterior distribution is,
\begin{adjustwidth}{-0.5cm}{}
\begin{eqnarray*}
\textnormal{log}P(\boldsymbol{\Theta}_{GMM} | \{ Y_i, \boldsymbol{M_i}, A_i, \boldsymbol{C_i}\}_{i=1}^n) &\propto& \sum_{i=1}^n \textnormal{log}P(Y_i | \boldsymbol{M_i}, \boldsymbol{\beta_m}, \sigma_e^2, \beta_a, \boldsymbol{\beta_c}, A_i, \boldsymbol{C_i}) + \textnormal{log}P(\boldsymbol{M_i} | \boldsymbol{\alpha_a}, \boldsymbol{\alpha_c}, \boldsymbol{\Sigma}, A_i, \boldsymbol{C_i})  \\ 
&& + \textnormal{log}P(\boldsymbol{\Theta}_{GMM}) \\
&=& \sum_{i=1}^n -\frac{1}{2}\textnormal{log}\sigma_e^2 - \frac{1}{2\sigma_e^2}(Y_i - \boldsymbol{M_i}^T\boldsymbol{\beta_m} - A_i\beta_a - \boldsymbol{C_i}^T\boldsymbol{\beta_c})^2 \nonumber \\
&-& \frac{1}{2}\textnormal{log}|\boldsymbol{\Sigma}| - \frac{1}{2}(\boldsymbol{M_i} -  A_i\boldsymbol{\alpha_a} - \boldsymbol{\alpha_c}\boldsymbol{C_i})^T \boldsymbol{\Sigma}^{-1}(\boldsymbol{M_i} -  A_i\boldsymbol{\alpha_a} - \boldsymbol{\alpha_c}\boldsymbol{C_i}) \nonumber \\
&+& \sum_{j=1}^p \sum_{k=1}^4 \gamma_{jk}(-\frac{d}{2}\textnormal{log} 2\pi - \frac{1}{2}\textnormal{log}|\boldsymbol{V_k}| - \frac{1}{2}\begin{bmatrix}
(\boldsymbol{\beta_m})_j \\
(\boldsymbol{\alpha_a})_j \end{bmatrix}^T\boldsymbol{V_k}^{-1}\begin{bmatrix}
(\boldsymbol{\beta_m})_j \\
(\boldsymbol{\alpha_a})_j \end{bmatrix}) \\
&-& \frac{q}{2}\textnormal{log}2\pi\sigma_c^2 - \frac{\boldsymbol{\beta_c}^T\boldsymbol{\beta_c}}{2\sigma_c^2} - \frac{pq}{2}\textnormal{log}2\pi\sigma_c^2 - \sum_{j=1}^p \frac{\boldsymbol{\alpha_{cj}}^T\boldsymbol{\alpha_{cj}}}{2\sigma_c^2} \\
&+& \sum_{j=1}^p \sum_{k=1}^4\gamma_{jk}\textnormal{log}(\pi_k) \\
&+& \sum_{k=1}^4a_k\textnormal{log}(\pi_k) + \sum_{k=1}^4(-\frac{\nu+d+1}{2}\textnormal{log}|\boldsymbol{V_k}| + \frac{1}{2}tr(\boldsymbol{\Psi_0}\boldsymbol{V_k}^{-1})))
\end{eqnarray*}
\end{adjustwidth}

\textit{Sampling $\begin{bmatrix}
(\boldsymbol{\beta_m})_j \\
(\boldsymbol{\alpha_a})_j \end{bmatrix}$ and $\gamma_{jk}$} \\
\begin{equation*}
    \textnormal{log}p(\begin{bmatrix}
(\boldsymbol{\beta_m})_j \\
(\boldsymbol{\alpha_a})_j \end{bmatrix} | \gamma_{jk} = 1, .) \propto -\frac{1}{2}\begin{bmatrix}
(\boldsymbol{\beta_m})_j \\
(\boldsymbol{\alpha_a})_j \end{bmatrix}^T (\boldsymbol{W_j} + \boldsymbol{V_k}^{-1})\begin{bmatrix}
(\boldsymbol{\beta_m})_j \\
(\boldsymbol{\alpha_a})_j \end{bmatrix} + \boldsymbol{w_j}^T\begin{bmatrix}
(\boldsymbol{\beta_m})_j \\
(\boldsymbol{\alpha_a})_j \end{bmatrix}
\end{equation*}
where $\boldsymbol{W_j} = \begin{bmatrix}
\sum_{i=1}^n (\sigma_e^2)^{-1} M_{ij}^2 & 0 \\
0 & \sum_{i=1}^n \color{black}{\boldsymbol{\Sigma}^{-1}A_i^2} \end{bmatrix}$ ($\boldsymbol{\Sigma}$ is diagonal, and can be replaced as $\sigma_g^2$), and $\boldsymbol{w_j} = (\sum_{i=1}^n (\sigma_e^2)^{-1}(Y_i - A_i\beta_a - \sum_{j^{'} \neq j}M_{ij^{'}}(\boldsymbol{\beta_m})_{j^{'}} )M_{ij}, \sum_{i=1}^n \boldsymbol{\Sigma}^{-1}M_{ij}A_i )^T$

\begin{equation*}
    p(\begin{bmatrix}
(\boldsymbol{\beta_m})_j \\
(\boldsymbol{\alpha_a})_j \end{bmatrix} | \gamma_{jk} = 1, .) \sim MVN_2((\boldsymbol{W_j} + \boldsymbol{V_k}^{-1})^{-1}\boldsymbol{w_j}, (\boldsymbol{W_j} + \boldsymbol{V_k}^{-1})^{-1}) 
\end{equation*}
\begin{equation*}
    \textnormal{log}p(\gamma_{jk} = 1|.) \propto -\frac{1}{2}\textnormal{log}|\boldsymbol{W_j}\boldsymbol{V_k} + \boldsymbol{I_2}| + \frac{1}{2}\boldsymbol{w_j}^T(\boldsymbol{W_j} + \boldsymbol{V_k}^{-1})^{-1}\boldsymbol{w_j} + \textnormal{log}(\pi_k)
\end{equation*}

\textit{Sampling $\pi_k$} \\
\begin{equation*}
    \{ \pi_{1}, \pi_{2}, \pi_{3}, \pi_{4} \} \propto \textnormal{Dirichlet}(a_{1}+\sum_{j=1}^p\gamma_{j1},a_{2}+\sum_{j=1}^p\gamma_{j2},a_{3}+\sum_{j=1}^p\gamma_{j3},a_{4}+\sum_{j=1}^p\gamma_{j4})
\end{equation*}

\textit{Sampling $\boldsymbol{V_k}$} \\
\begin{equation*}
    \textnormal{log}p(\boldsymbol{V_k}|.) \propto -\frac{1}{2}(\sum_{j=1}^p \gamma_{jk}+\nu+d+1)\textnormal{log}|\boldsymbol{V_k}| - \frac{1}{2}tr(\boldsymbol{\boldsymbol{\Psi_0}}\boldsymbol{V_k}^{-1}) + \sum_{j=1}^p \gamma_{jk}(-\frac{1}{2}\begin{bmatrix}
(\boldsymbol{\beta_m})_j \\
(\boldsymbol{\alpha_a})_j \end{bmatrix}^T\boldsymbol{V_k}^{-1}\begin{bmatrix}
(\boldsymbol{\beta_m})_j \\
(\boldsymbol{\alpha_a})_j \end{bmatrix}) 
\end{equation*}
\begin{equation*}
    p(\boldsymbol{V_k}|.) \sim \textnormal{Inv-Wishart} (\boldsymbol{\Psi_0 + \sum_{j=1}^p \gamma_{jk} \begin{bmatrix}
(\boldsymbol{\beta_m})_j \\
(\boldsymbol{\alpha_a})_j \end{bmatrix}\begin{bmatrix}
(\boldsymbol{\beta_m})_j \\
(\boldsymbol{\alpha_a})_j \end{bmatrix}}^T, \sum_{j=1}^p \gamma_{jk}+\nu)
\end{equation*}

\textit{Sampling $\beta_a$}
\begin{equation*}
    \textnormal{log}p(\beta_{a}| .) \propto  -\frac{\beta_{a}^2}{2\sigma_a^2} - \sum_{i=1}^n \{ \frac{(A_i\beta_{a})^2}{2\sigma_e^2} - \sigma_1^{-2}A_{i}(Y_i - \boldsymbol{M_i}^T\boldsymbol{\beta_m} - \boldsymbol{C_i}^T\boldsymbol{\beta_c})\beta_{a} \}
\end{equation*}
\begin{equation*}
    p(\beta_{a}| .) \sim N(\frac{\sum_{i=1}^n A_{i}(Y_i - \boldsymbol{M_i}^T\boldsymbol{\beta_m} - \boldsymbol{C_i}^T\boldsymbol{\beta_c})}{\sigma_e^2/\sigma_a^2 + \sum_{i=1}^n A_{i}^2}, \frac{1}{1/\sigma_a^2 + \sum_{i=1}^n A_{i}^2/\sigma_e^2})
\end{equation*}

\textit{Sampling $\sigma_a^2$} \\
\begin{equation*}
    \textnormal{log}p(\sigma_a^2|.) \propto -(\frac{1}{2} + h_a + 1)\textnormal{log}(\sigma_a^2) - (\frac{\beta_{a}^2}{2}+l_a)\sigma_a^{-2}
\end{equation*}
\begin{equation*}
    p(\sigma_a^2|.) \sim \textnormal{inverse-gamma}(\frac{1}{2} + h_a, \frac{\beta_{a}^2}{2}+l_a)
\end{equation*}

\textit{Sampling $\sigma_e^2$} \\
\begin{equation*}
    \textnormal{log}p(\sigma_e^2|.) = -(\frac{n}{2} + h_1 + 1)\textnormal{log}(\sigma_e^2) - (\frac{\sum_{i=1}^n (Y_i - \boldsymbol{M_i}^T\boldsymbol{\beta_m} - A_i\beta_a - \boldsymbol{C_i}^T\boldsymbol{\beta_c})^2}{2} +l_1)\sigma_1^{-2}
\end{equation*}
\begin{equation*}
    p(\sigma_e^2|.) \sim \textnormal{inverse-gamma}(\frac{n}{2} + h_1, \frac{\sum_{i=1}^n (Y_i - \boldsymbol{M_i}^T\boldsymbol{\beta_m} - A_i\beta_a - \boldsymbol{C_i}^T\boldsymbol{\beta_c})^2}{2}+l_1)
\end{equation*}

\textit{Sampling $\sigma_g^2$} \\
\begin{equation*}
    \textnormal{log}p(\sigma_g^2|.) = -(\frac{pn}{2} + h_2 + 1)\textnormal{log}(\sigma_g^2) - (\frac{\sum_{i=1}^n (\boldsymbol{M_i}^T - A_i\boldsymbol{\alpha_a} - \boldsymbol{C_i}^T\boldsymbol{\alpha_c})(\boldsymbol{M_i}^T - A_i\boldsymbol{\alpha_a} - \boldsymbol{C_i}^T\boldsymbol{\alpha_c})^T}{2} +l_2)\sigma_2^{-2}
\end{equation*}
\begin{equation*}
    p(\sigma_g^2|.) \sim \textnormal{inverse-gamma}(\frac{pn}{2} + h_2, \frac{\sum_{i=1}^n (\boldsymbol{M_i}^T - A_i\boldsymbol{\alpha_a} - \boldsymbol{C_i}^T\boldsymbol{\alpha_c})(\boldsymbol{M_i}^T - A_i\boldsymbol{\alpha_a} - \boldsymbol{C_i}^T\boldsymbol{\alpha_c})^T}{2} +l_2)
\end{equation*}

\textit{Sampling $\beta_{cw}$} \\
\begin{equation*}
    \textnormal{log}p(\beta_{cw}| .) = - \sum_{i=1}^n \{ \frac{(C_{iw}\beta_{cw})^2}{2\sigma_e^2} + \sigma_e^{-2}C_{iw}(Y_i - \boldsymbol{M_i}^T\boldsymbol{\beta_m} - A_i\beta_a - \sum_{s \neq w}C_{iw}\beta_{cw})\beta_{cw} \}
\end{equation*}
\begin{equation*}
    p(\beta_{cw}|.) = N( \frac{\sum_{i=1}^nC_{iw}(Y_i - A_i\beta_a - \boldsymbol{M_i}^T\boldsymbol{\beta_m} - \sum_{s \neq w}C_{iw}\beta_{cw})}{\sum_{i=1}^n C_{iw}^2}, \frac{\sigma_e^2}{\sum_{i=1}^n C_{iw}^2})  \\
\end{equation*}

\textit{Sampling $(\boldsymbol{\alpha_{cw}})_j$} \\
\begin{equation*}
    \textnormal{log}p((\boldsymbol{\alpha_{cw}})_j| .) = - \sum_{i=1}^n \{ \frac{(C_{iw}(\boldsymbol{\alpha_{cw}})_j)^2}{2\sigma_g^2} + \sigma_g^{-2}C_{iw} (M_i^{(j)} - A_i\alpha_{aj} - \sum_{s \neq w}C_{is}(\boldsymbol{\alpha_{cs}})_j)(\boldsymbol{\alpha_{cw}})_j \}
\end{equation*}
\begin{equation*}
    p((\boldsymbol{\alpha_{cw}})_j|.) = N(\frac{\sum_{i=1}^nC_{iw} (M_i^{(j)} - A_i\alpha_{aj} - \sum_{s \neq w}C_{is}(\boldsymbol{\alpha_{cs}})_j)}{\sum_{i=1}^n C_{iw}^2}, \frac{\sigma_g^2}{\sum_{i=1}^n C_{iw}^2}) \\
\end{equation*}

\section{Posterior Sampling Algorithm Details for Product Threshold Gaussian (PTG) Prior}
\hfill \break
Let $\boldsymbol{\Theta}_{PTG} = (\boldsymbol{\beta_m}, \boldsymbol{\alpha_a}, \boldsymbol{\tilde{\beta}_m},  \boldsymbol{\tilde{\alpha}_a}, \tau_{\beta}^2, \tau_{\alpha}^2, \beta_a, \boldsymbol{\beta_c}, \boldsymbol{\alpha_c}, \sigma_e^2, \boldsymbol{\Sigma})$ denote all the unknown parameters in the model. Under the PTG prior, the joint log posterior distribution is,
\vspace{-0.3in}
\begin{adjustwidth}{-0.5cm}{}
\begin{eqnarray*}
\log P(\boldsymbol{\Theta}_{PTG}| \{ Y_i, \boldsymbol{M_i}, A_i, \boldsymbol{C_i}\}_{i=1}^n) &\propto& \sum_{i=1}^n \log P(Y_i | \boldsymbol{M_i}, \boldsymbol{\beta_m}, \sigma_e^2, \beta_a, \boldsymbol{\beta_c}, A_i, \boldsymbol{C_i}) + \log P(\boldsymbol{M_i} | \boldsymbol{\alpha_a}, \boldsymbol{\alpha_c}, \boldsymbol{\Sigma}, A_i, \boldsymbol{C_i})  \\ 
&& + \log P(\boldsymbol{\Theta}_{PTG}) \\
&=& \sum_{i=1}^n -\frac{1}{2}\log \sigma_e^2 - \frac{1}{2\sigma_e^2}(Y_i - \boldsymbol{M_i}^T\boldsymbol{\beta_m} - A_i\beta_a - \boldsymbol{C_i}^T\boldsymbol{\beta_c})^2 \nonumber \\
&-& \frac{1}{2}\log |\boldsymbol{\Sigma}| - \frac{1}{2}(\boldsymbol{M_i} -  A_i\boldsymbol{\alpha_a} - \boldsymbol{\alpha_c}\boldsymbol{C_i})^T \boldsymbol{\Sigma}^{-1}(\boldsymbol{M_i} -  A_i\boldsymbol{\alpha_a} - \boldsymbol{\alpha_c}\boldsymbol{C_i}) \nonumber \\
&+& \sum_{i=1}^p -\frac{1}{2}\log \tau_{\beta}^2 - \frac{(\boldsymbol{\tilde{\beta}_m})_j^2}{2\tau_{\beta}^2} + \sum_{i=1}^p -\frac{1}{2}\log \tau_{\alpha}^2 - \frac{(\boldsymbol{\tilde{\alpha}_a})_j^2}{2\tau_{\alpha}^2} \\
&-& \frac{q}{2}\log 2\pi\sigma_c^2 - \frac{\boldsymbol{\beta_c}^T\boldsymbol{\beta_c}}{2\sigma_c^2} - \frac{pq}{2}\log 2\pi\sigma_c^2 - \sum_{j=1}^p \frac{\boldsymbol{\alpha_{cj}}^T\boldsymbol{\alpha_{cj}}}{2\sigma_c^2} 
\end{eqnarray*}
\end{adjustwidth}

\textit{Sampling $(\boldsymbol{\beta_m})_j$} \\
For $(\boldsymbol{\tilde{\beta}_m})_j$, we denote its threshold conditional on the other parameters as 
\[
    u_{(\boldsymbol{\tilde{\beta}_m})_j} = \Big\{\begin{array}{lr}
        \textnormal{min}(\lambda_1, \lambda_0/|(\boldsymbol{\tilde{\alpha}_a})_j|), & \text{for } (\boldsymbol{\tilde{\alpha}_a})_j \neq 0 \\
        \lambda_1, & \text{for } (\boldsymbol{\tilde{\alpha}_a})_j = 0 \\
        \end{array} 
  \]
\begin{equation*}
     \textnormal{log}p((\boldsymbol{\tilde{\beta}_m})_j|             |(\boldsymbol{\tilde{\beta}_m})_j| < u_{(\boldsymbol{\tilde{\beta}_m})_j}) \propto -(\boldsymbol{\tilde{\beta}_m})_j^2/(2\tau_{\beta}^{2}) 
\end{equation*}
\begin{equation*}
     (\boldsymbol{\tilde{\beta}_m})_j|  |(\boldsymbol{\tilde{\beta}_m})_j| < u_{(\boldsymbol{\tilde{\beta}_m})_j} \sim TN(0, \tau_{\beta}^{2}, -u_{(\boldsymbol{\tilde{\beta}_m})_j}, u_{(\boldsymbol{\tilde{\beta}_m})_j})
\end{equation*}
where $TN(\mu, \sigma^2, a, b)$ denotes a truncated normal distribution with mean $\mu$, variance $\sigma^2$ truncated between $[a, b]$.
\begin{align*}
   \lefteqn{ \textnormal{log}p((\boldsymbol{\tilde{\beta}_m})_j|             |(\boldsymbol{\tilde{\beta}_m})_j| >= u_{(\boldsymbol{\tilde{\beta}_m})_j})}\\
   &\propto -\frac{(\boldsymbol{\tilde{\beta}_m})_j^2}{2\tau_{\beta}^2} - \sum_{i=1}^n \{ \frac{(M_i^{(j)}(\boldsymbol{\tilde{\beta}_m})_j)^2}{2\sigma_e^2} + \sigma_e^{-2}M_i^{(j)}(Y_i - A_i\beta_a - \sum_{s \neq j}M_i^{(s)}(\boldsymbol{\tilde{\beta}_m})_s - \boldsymbol{C_i}^T\boldsymbol{\beta_c})(\boldsymbol{\tilde{\beta}_m})_j \}
\end{align*}
\begin{eqnarray*}
    (\boldsymbol{\tilde{\beta}_m})_j|             (\boldsymbol{\tilde{\beta}_m})_j >= u_{(\boldsymbol{\tilde{\beta}_m})_j} \sim TN(\mu_{mj}, s_{mj}^2, u_{(\boldsymbol{\tilde{\beta}_m})_j}, \infty) \\
    (\boldsymbol{\tilde{\beta}_m})_j|             (\boldsymbol{\tilde{\beta}_m})_j <= -u_{(\boldsymbol{\tilde{\beta}_m})_j} \sim TN(\mu_{mj}, s_{mj}^2, -\infty, -u_{(\boldsymbol{\tilde{\beta}_m})_j})
\end{eqnarray*}
\begin{equation*}
    \mu_{mj} = \frac{\sum_{i=1}^nM_i^{(j)}(Y_i - A_i\beta_a - \sum_{s \neq j}M_i^{(s)}(\boldsymbol{\tilde{\beta}_m})_s - \boldsymbol{C_i}^T\boldsymbol{\beta_c})}{\sigma_e^2/\tau_{\beta}^2 + \sum_{i=1}^n (M_i^{(j)})^2}, 
    s_{mj}^2 = \frac{1}{1/\tau_{\beta}^2 + \sum_{i=1}^n (M_i^{(j)})^2/\sigma_e^2} \\
\end{equation*}
And,
\begin{eqnarray*}
    p(|(\boldsymbol{\tilde{\beta}_m})_j| < u_{(\boldsymbol{\tilde{\beta}_m})_j}) = \frac{B_1}{B_1 + B_2 + B_3} \\
    p((\boldsymbol{\tilde{\beta}_m})_j >= u_{(\boldsymbol{\tilde{\beta}_m})_j}) = \frac{B_2}{B_1 + B_2 + B_3} \\
    p((\boldsymbol{\tilde{\beta}_m})_j <= -u_{(\boldsymbol{\tilde{\beta}_m})_j}) = \frac{B_3}{B_1 + B_2 + B_3}
\end{eqnarray*}
where $B_1 = \int_{-u_{(\boldsymbol{\tilde{\beta}_m})_j}}^{u_{(\boldsymbol{\tilde{\beta}_m})_j}}\frac{1}{\sqrt{2\pi\tau_{\beta}^2}}\textnormal{exp}(-\frac{(\boldsymbol{\tilde{\beta}_m})_j^2}{2\tau_{\beta}^2}) = 1 - 2\Phi(-\frac{u_{(\boldsymbol{\tilde{\beta}_m})_j}}{\tau_{\beta}^2})$, $\Phi(x)$ is the CDF for standard normal distribution, $B_2 = \textnormal{exp}(\mu_{mj}^2/2s_{mj}^2 +  \textnormal{log}(s_{mj}) - \textnormal{log}(\tau_{\beta}))(1-\Phi(\frac{u_{(\boldsymbol{\tilde{\beta}_m})_j}}{\tau_{\beta}^2}))$, $B_3 = \textnormal{exp}(\mu_{mj}^2/2s_{mj}^2 +  \textnormal{log}(s_{mj}) - \textnormal{log}(\tau_{\beta}))\Phi(-\frac{u_{(\boldsymbol{\tilde{\beta}_m})_j}}{\tau_{\beta}^2})$.
\[
    (\boldsymbol{\beta_m})_j = \Big\{\begin{array}{lr}
       (\boldsymbol{\tilde{\beta}_m})_j , & \text{for } |(\boldsymbol{\tilde{\beta}_m})_j| >= u_{(\boldsymbol{\tilde{\beta}_m})_j} \\
        0 , & \text{for } |(\boldsymbol{\tilde{\beta}_m})_j| < u_{(\boldsymbol{\tilde{\beta}_m})_j} \\
        \end{array} 
 \]
\textit{Sampling $(\boldsymbol{\alpha_a})_j$} \\
For $(\boldsymbol{\tilde{\alpha}_a})_j$, we denote its threshold conditional on the other parameters as 
\[
    u_{(\boldsymbol{\tilde{\alpha}_a})_j} = \Big\{\begin{array}{lr}
        \textnormal{min}(\lambda_2, \lambda_0/|(\boldsymbol{\tilde{\beta}_m})_j|), & \text{for } (\boldsymbol{\tilde{\beta}_m})_j \neq 0 \\
        \lambda_2, & \text{for } (\boldsymbol{\tilde{\beta}_m})_j = 0 \\
        \end{array} 
  \]
\begin{equation*}
     \textnormal{log}p((\boldsymbol{\tilde{\alpha}_a})_j|             |(\boldsymbol{\tilde{\alpha}_a})_j| < u_{(\boldsymbol{\tilde{\alpha}_a})_j}) \propto -(\boldsymbol{\tilde{\alpha}_a})_j^2/(2\tau_{\alpha}^{2}) 
\end{equation*}
\begin{equation*}
     (\boldsymbol{\tilde{\alpha}_a})_j|  |(\boldsymbol{\tilde{\alpha}_a})_j| < u_{(\boldsymbol{\tilde{\alpha}_a})_j} \sim TN(0, \tau_{\alpha}^{2}, -u_{(\boldsymbol{\tilde{\alpha}_a})_j}, u_{(\boldsymbol{\tilde{\alpha}_a})_j})
\end{equation*}
\begin{equation*}
    \textnormal{log}p((\boldsymbol{\tilde{\alpha}_a})_j|             |(\boldsymbol{\tilde{\alpha}_a})_j| >= u_{(\boldsymbol{\tilde{\alpha}_a})_j}) \propto -\frac{(\boldsymbol{\tilde{\alpha}_a})_j^2}{2\tau_{\alpha}^2} - \sum_{i=1}^n \{ \frac{(A_i(\boldsymbol{\tilde{\alpha}_a})_j)^2}{2\sigma_g^2} + \sigma_g^{-2}A_i(M_i^{(j)} - (\boldsymbol{\alpha_c}\boldsymbol{C_i})_j)(\boldsymbol{\tilde{\alpha}_a})_j \}
\end{equation*}
\begin{eqnarray*}
    (\boldsymbol{\tilde{\alpha}_a})_j|             (\boldsymbol{\tilde{\alpha}_a})_j >= u_{(\boldsymbol{\tilde{\alpha}_a})_j} \sim TN(\mu_{aj}, s_{aj}^2, u_{(\boldsymbol{\tilde{\alpha}_a})_j}, \infty) \\
    (\boldsymbol{\tilde{\alpha}_a})_j|             (\boldsymbol{\tilde{\alpha}_a})_j <= -u_{(\boldsymbol{\tilde{\alpha}_a})_j} \sim TN(\mu_{aj}, s_{aj}^2, -\infty, -u_{(\boldsymbol{\tilde{\alpha}_a})_j})
\end{eqnarray*}
\begin{equation*}
    \mu_{aj} = \frac{\sum_{i=1}^n A_i(M_i^{(j)}- (\boldsymbol{\alpha_c}\boldsymbol{C_i})_j)}{\sigma_g^2/\tau_{\alpha}^2 + \sum_{i=1}^n A_i^2}, 
    s_{aj}^2 =  \frac{1}{1/\tau_{\alpha}^2 + \sum_{i=1}^n A_i^2/\sigma_g^2}\\
\end{equation*}
And,
\begin{eqnarray*}
    p(|(\boldsymbol{\tilde{\alpha}_a})_j| < u_{(\boldsymbol{\tilde{\alpha}_a})_j}) = \frac{A_1}{A_1 + A_2 + A_3} \\
    p((\boldsymbol{\tilde{\alpha}_a})_j >= u_{(\boldsymbol{\tilde{\alpha}_a})_j}) = \frac{A_2}{A_1 + A_2 + A_3} \\
    p((\boldsymbol{\tilde{\alpha}_a})_j <= -u_{(\boldsymbol{\tilde{\alpha}_a})_j}) = \frac{A_3}{A_1 + A_2 + A_3}
\end{eqnarray*}
where $A_1 = \int_{-u_{(\boldsymbol{\tilde{\alpha}_a})_j}}^{u_{(\boldsymbol{\tilde{\alpha}_a})_j}}\frac{1}{\sqrt{2\pi\tau_{\alpha}^2}}\textnormal{exp}(-\frac{(\boldsymbol{\tilde{\alpha}_a})_j^2}{2\tau_{\alpha}^2}) = 1 - 2\Phi(-\frac{u_{(\boldsymbol{\tilde{\alpha}_a})_j}}{\tau_{\alpha}^2})$, $\Phi(x)$ is the CDF for standard normal distribution, $A_2 = \textnormal{exp}(\mu_{aj}^2/2s_{aj}^2 +  \textnormal{log}(s_{aj}) - \textnormal{log}(\tau_{\alpha}))(1-\Phi(\frac{u_{(\boldsymbol{\tilde{\alpha}_a})_j}}{\tau_{\alpha}^2}))$, $A_3 = \textnormal{exp}(\mu_{aj}^2/2s_{aj}^2 +  \textnormal{log}(s_{aj}) - \textnormal{log}(\tau_{\alpha}))\Phi(-\frac{u_{(\boldsymbol{\tilde{\alpha}_a})_j}}{\tau_{\alpha}^2})$.
\[
    (\boldsymbol{\alpha_a})_j = \Big\{\begin{array}{lr}
       (\boldsymbol{\tilde{\alpha}_a})_j , & \text{for } |(\boldsymbol{\tilde{\alpha}_a})_j| >= u_{(\boldsymbol{\tilde{\alpha}_a})_j} \\
        0 , & \text{for } |(\boldsymbol{\tilde{\alpha}_a})_j| < u_{(\boldsymbol{\tilde{\alpha}_a})_j} \\
        \end{array} 
 \]

\textit{Sampling $\beta_a$}
\begin{equation*}
    \textnormal{log}p(\beta_{a}| .) \propto  -\frac{\beta_{a}^2}{2\sigma_a^2} - \sum_{i=1}^n \{ \frac{(A_i\beta_{a})^2}{2\sigma_1^2} - \sigma_1^{-2}A_{i}(Y_i - \boldsymbol{M_i}^T\boldsymbol{\beta_m} - \boldsymbol{C_i}^T\boldsymbol{\beta_c})\beta_{a} \}
\end{equation*}
\begin{equation*}
    p(\beta_{a}| .) \sim N(\frac{\sum_{i=1}^n A_{i}(Y_i - \boldsymbol{M_i}^T\boldsymbol{\beta_m} - \boldsymbol{C_i}^T\boldsymbol{\beta_c})}{\sigma_1^2/\sigma_a^2 + \sum_{i=1}^n A_{i}^2}, \frac{1}{1/\sigma_a^2 + \sum_{i=1}^n A_{i}^2/\sigma_1^2})
\end{equation*}

\textit{Sampling $\sigma_a^2$} \\
\begin{equation*}
    \textnormal{log}p(\sigma_a^2|.) \propto -(\frac{1}{2} + h_a + 1)\textnormal{log}(\sigma_a^2) - (\frac{\beta_{a}^2}{2}+l_a)\sigma_a^{-2}
\end{equation*}
\begin{equation*}
    p(\sigma_a^2|.) \sim \textnormal{inverse-gamma}(\frac{1}{2} + h_a, \frac{\beta_{a}^2}{2}+l_a)
\end{equation*}

\textit{Sampling $\sigma_e^2$} \\
\begin{equation*}
    \textnormal{log}p(\sigma_e^2|.) = -(\frac{n}{2} + h_1 + 1)\textnormal{log}(\sigma_e^2) - (\frac{\sum_{i=1}^n (Y_i - \boldsymbol{M_i}^T\boldsymbol{\beta_m} - A_i\beta_a - \boldsymbol{C_i}^T\boldsymbol{\beta_c})^2}{2} +l_1)\sigma_e^{-2}
\end{equation*}
\begin{equation*}
    p(\sigma_e^2|.) \sim \textnormal{inverse-gamma}(\frac{n}{2} + h_1, \frac{\sum_{i=1}^n (Y_i - \boldsymbol{M_i}^T\boldsymbol{\beta_m} - A_i\beta_a - \boldsymbol{C_i}^T\boldsymbol{\beta_c})^2}{2}+l_1)
\end{equation*}

\textit{Sampling $\sigma_g^2$} \\
\begin{equation*}
    \textnormal{log}p(\sigma_g^2|.) = -(\frac{pn}{2} + h_2 + 1)\textnormal{log}(\sigma_g^2) - (\frac{\sum_{i=1}^n (\boldsymbol{M_i}^T - A_i\boldsymbol{\alpha_a} - \boldsymbol{C_i}^T\boldsymbol{\alpha_c})(\boldsymbol{M_i}^T - A_i\boldsymbol{\alpha_a} - \boldsymbol{C_i}^T\boldsymbol{\alpha_c})^T}{2} +l_2)\sigma_g^{-2}
\end{equation*}
\begin{equation*}
    p(\sigma_g^2|.) \sim \textnormal{inverse-gamma}(\frac{pn}{2} + h_2, \frac{\sum_{i=1}^n (\boldsymbol{M_i}^T - A_i\boldsymbol{\alpha_a} - \boldsymbol{C_i}^T\boldsymbol{\alpha_c})(\boldsymbol{M_i}^T - A_i\boldsymbol{\alpha_a} - \boldsymbol{C_i}^T\boldsymbol{\alpha_c})^T}{2} +l_2)
\end{equation*}

\textit{Sampling $\tau_{\beta}^2$} \\
\begin{equation*}
    \textnormal{log}p(\tau_{\beta}^2|.) = -(\frac{q}{2} + k_m + 1)\textnormal{log}(\tau_{\beta}^2) - (\frac{\sum_{j=1}^q (\boldsymbol{\tilde{\beta}_m})_j^2}{2}+l_m)\tau_{\beta}^{-2}
\end{equation*}
\begin{equation*}
    p(\tau_{\beta}^2|.) \sim \textnormal{inverse-gamma}(\frac{q}{2} + k_m, \frac{\sum_{j=1}^q (\boldsymbol{\tilde{\beta}_m})_j^2}{2}+l_m)
\end{equation*}

\textit{Sampling $\tau_{\alpha}^2$} \\
\begin{equation*}
    \textnormal{log}p(\tau_{\alpha}^2|.) = -(\frac{q}{2} + k_{ma} + 1)\textnormal{log}(\tau_{\alpha}^2) - (\frac{\sum_{j=1}^q (\boldsymbol{\tilde{\alpha}_a})_j^2}{2}+l_{ma})\tau_{\alpha}^{-2}
\end{equation*}
\begin{equation*}
    p(\tau_{\alpha}^2|.) \sim \textnormal{inverse-gamma}(\frac{q}{2} + k_{ma}, \frac{\sum_{j=1}^q (\boldsymbol{\tilde{\alpha}_a})_j^2}{2}+l_{ma})
\end{equation*}

\textit{Sampling $\beta_{cw}$} \\
\begin{equation*}
    \textnormal{log}p(\beta_{cw}| .) = - \sum_{i=1}^n \{ \frac{(C_{iw}\beta_{cw})^2}{2\sigma_e^2} + \sigma_e^{-2}C_{iw}(Y_i - \boldsymbol{M_i}^T\boldsymbol{\beta_m} - A_i\beta_a - \sum_{s \neq w}C_{iw}\beta_{cw})\beta_{cw} \}
\end{equation*}
\begin{equation*}
    p(\beta_{cw}|.) = N( \frac{\sum_{i=1}^nC_{iw}(Y_i - A_i\beta_a - \boldsymbol{M_i}^T\boldsymbol{\beta_m} - \sum_{s \neq w}C_{iw}\beta_{cw})}{\sum_{i=1}^n C_{iw}^2}, \frac{\sigma_e^2}{\sum_{i=1}^n C_{iw}^2})  \\
\end{equation*}

\textit{Sampling $(\boldsymbol{\alpha_{cw}})_j$} \\
\begin{equation*}
    \textnormal{log}p((\boldsymbol{\alpha_{cw}})_j| .) = - \sum_{i=1}^n \{ \frac{(C_{iw}(\boldsymbol{\alpha_{cw}})_j)^2}{2\sigma_g^2} + \sigma_g^{-2}C_{iw} (M_i^{(j)} - A_i\alpha_{aj} - \sum_{s \neq w}C_{is}(\boldsymbol{\alpha_{cs}})_j)(\boldsymbol{\alpha_{cw}})_j \}
\end{equation*}
\begin{equation*}
    p((\boldsymbol{\alpha_{cw}})_j|.) = N(\frac{\sum_{i=1}^nC_{iw} (M_i^{(j)} - A_i\alpha_{aj} - \sum_{s \neq w}C_{is}(\boldsymbol{\alpha_{cs}})_j)}{\sum_{i=1}^n C_{iw}^2}, \frac{\sigma_g^2}{\sum_{i=1}^n C_{iw}^2}) \\
\end{equation*}

\section{Effects Distribution, Empirical FDR Results and Computing Time in Simulations} 
\begin{figure}
\begin{center}
\includegraphics[scale=0.30]{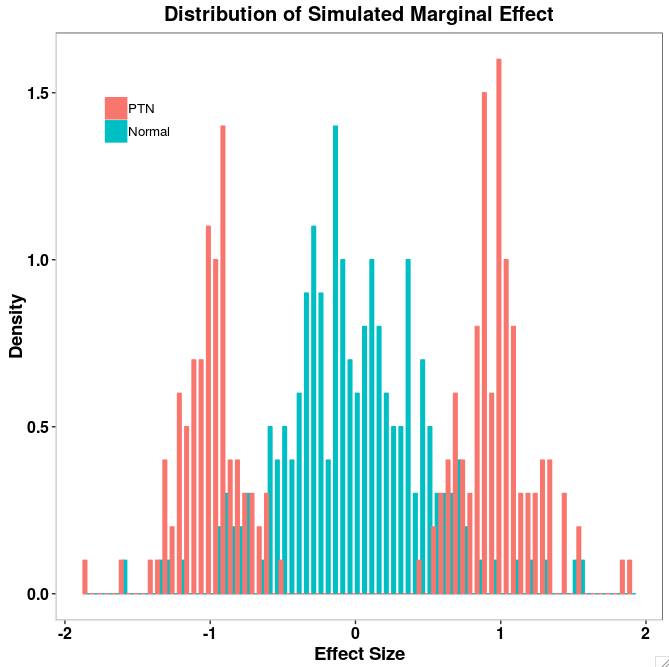}
\includegraphics[scale=0.30]{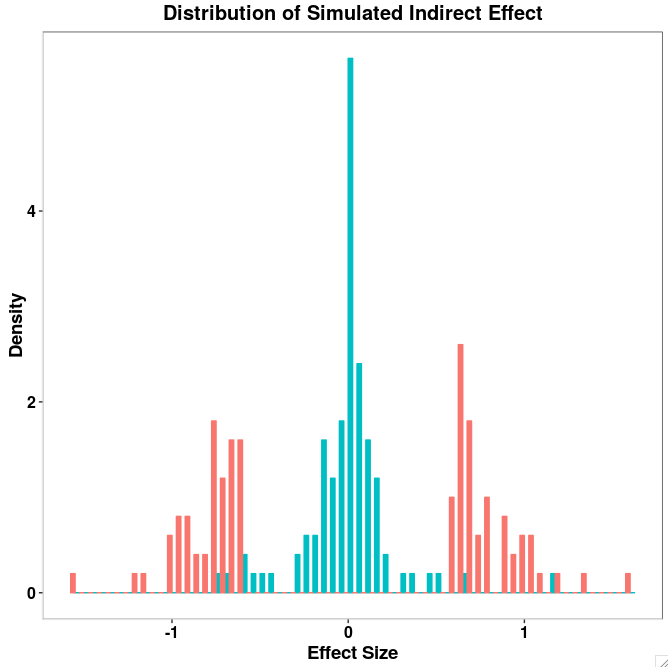}
\includegraphics[scale=0.30]{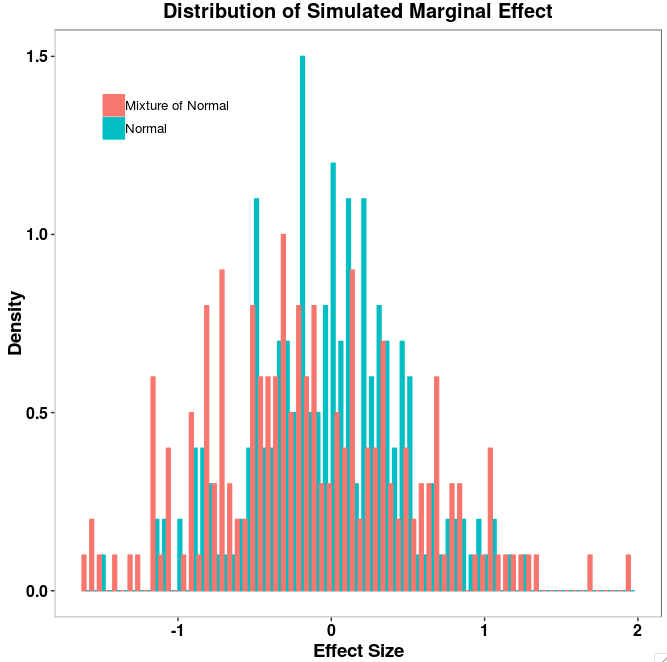}
\includegraphics[scale=0.30]{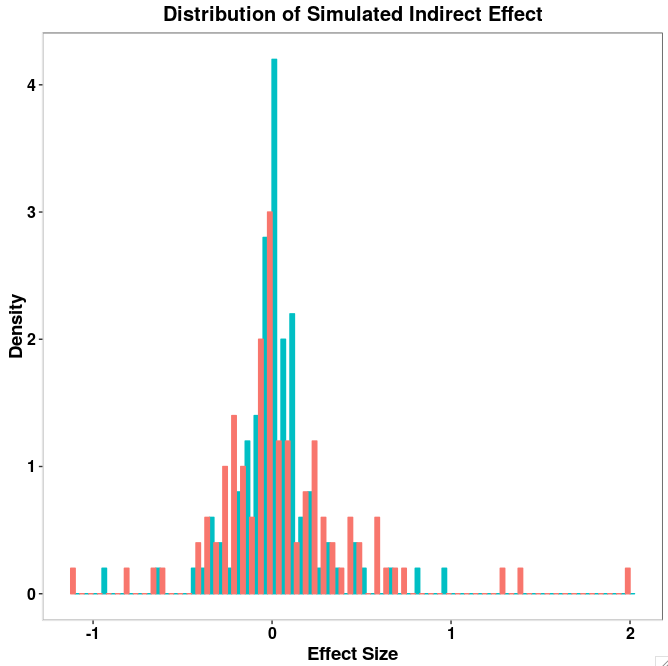}
\includegraphics[scale=0.30]{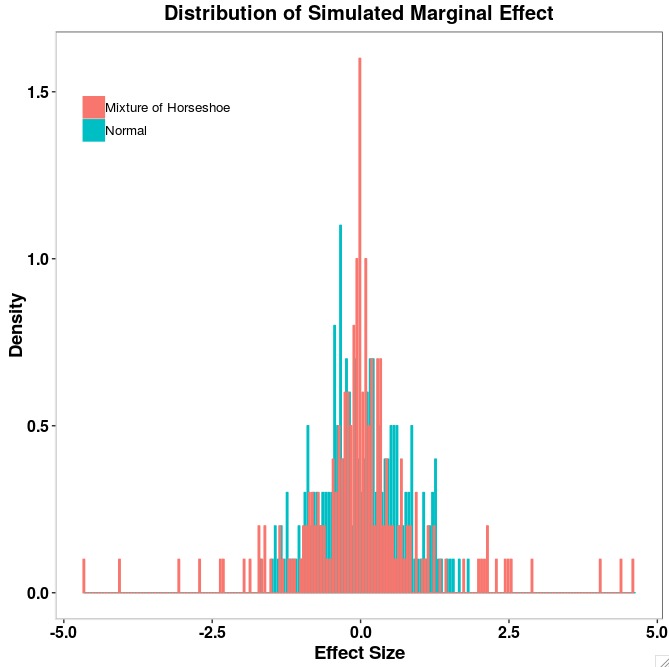}
\includegraphics[scale=0.30]{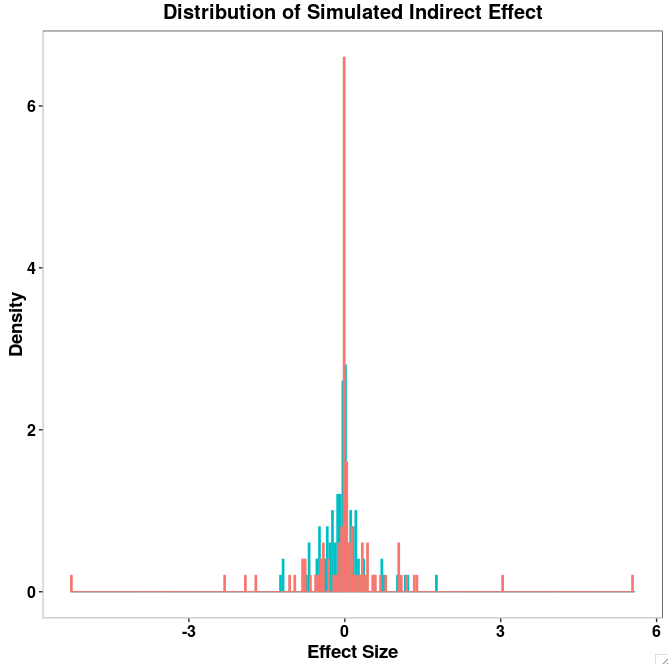}
\captionof{figure}{The distributions of the simulated non-zero marginal effects, $(\boldsymbol{\beta_m})_j$ (or $(\boldsymbol{\alpha_a})_j$) and indirect effects, $(\boldsymbol{\beta_m})_j(\boldsymbol{\alpha_a})_j$ under the three simulation settings when $n = 100, p = 200$. Each row represents one scenario, i.e. effects under prior model PTG, GMM and Mixture of Horseshoe. We include marginal effects from normals with the same variances as the simulation distributions and the corresponding indirect effects as a comparison.}
\label{fig:setting}
\end{center}
\end{figure}

\hfill \break
To better understand the generated effects under the three different data generating mechanism in the simulation Setting (A)-(C), we examine the corresponding distributions of the simulated non-zero marginal effects, $(\boldsymbol{\beta_m})_j$ (or $(\boldsymbol{\alpha_a})_j$) and indirect effects, $(\boldsymbol{\beta_m})_j(\boldsymbol{\alpha_a})_j$ in Figure \ref{fig:setting}.

The PTG prior model essentially produces effects truncated away from zero (Setting (A)), where the thresholding parameter $\lambda=\left(\lambda_{0}, \lambda_{1}, \lambda_{2}\right)$ is determined by the proportion of non-zero effects. For example, choosing $\lambda_0 = |\boldsymbol{\tilde{\alpha}_a}\boldsymbol{\tilde{\beta}_m}|^{(95)}, \lambda_1 = |\boldsymbol{\tilde{\beta}_m}|^{(85)}, \lambda_2 = |\boldsymbol{\tilde{\alpha}_a}|^{(93)}$ approximately makes $\pi_{1} = 0.05, \pi_{2} = 0.10, \pi_{3} = 0.05, \pi_{4} = 0.80$. The relatively small non-zero marginal effects are picked up by its indirect effects exceeding the product threshold. The Setting (B) with four components of bivariate Gaussian mixture is straightforward, and the resulting indirect effects distribute as a product of two normal distributions. Under the Setting (C), we can see that the horseshoe distribution has a tall spike near zero and heavy tails on large effects, and this generates uneven effects different from either PTG or GMM prior model. The distribution of the corresponding indirect effects show a stronger contrast between small and large effects. 

As a practical procedure, we suggest a cutoff on the posterior inclusion probabilities (PIP) to identify a significance threshold for declaring active mediators. To evaluate the performance of this significance rule, we report the empirical FDR and TPR in Table \ref{tbl:efdr1} and \ref{tbl:efdr2} under all the simulation scenarios. We find that at PIP = 0.5 cutoff, the two proposed methods, PTG and GMM, exhibit good selection performance while maintaining a reasonable FDR in most scenarios. At PIP = 0.9 cutoff, the two methods provide over conservative estimates of FDR, leading to reduced power in mediator selection. Therefore, we will use the 0.5 cutoff on the PIPs as a selection criterion in the following applications.

\begin{table}[!h]
\begin{adjustwidth}{0.0cm}{}
\scalebox{0.9}{
\centering{
  \begin{tabular}{c  c c c c c}
    \hline     
    Method & TPR(FDR=0.1) & TPR(PIP$>$0.9) & FDR(PIP$>$0.9) & TPR(PIP$>$0.5) & FDR(PIP$>$0.5)\\
    \hline
    \multicolumn{6}{c}{$n = 100, p = 200, p_{11} = 10, \textit{fixed effects (I)}$} \\
    \hline
    PTG  & 0.54(0.025) & 0.27(0.017) & 0.03(0.014) & 0.55(0.017) & 0.13(0.016) \\
    GMM & 0.42(0.023) & 0.17(0.022) & 0.03(0.021) & 0.44(0.023) & 0.16(0.017) \\
    \hline
    \multicolumn{6}{c}{$n = 100, p = 200, p_{11} = 10, \textit{fixed effects (II)}$} \\
    \hline
    PTG  & 0.34(0.017) & 0.27(0.008) & 0.04(0.019) & 0.37(0.013) & 0.14(0.019) \\
    GMM & 0.39(0.020) & 0.21(0.010) & 0.03(0.016) & 0.39(0.013) & 0.11(0.017) \\
    \hline
    \multicolumn{6}{c}{$n = 100, p = 200, p_{11} = 10, PTG$} \\
    \hline
    PTG  & 0.45(0.020) & 0.19(0.014) & 0.01(0.007) & 0.49(0.018) & 0.18(0.015) \\
    GMM & 0.43(0.015) & 0.26(0.011) & 0.03(0.012) & 0.45(0.014) & 0.11(0.012) \\
	\hline
	\multicolumn{6}{c}{$n = 100, p = 200, p_{11} = 10, Gaussian$} \\
    \hline
    PTG  & 0.38(0.008) & 0.26(0.008) & 0.01(0.006) & 0.56(0.010) & 0.39(0.011) \\
    GMM & 0.41(0.006) & 0.27(0.005) & 0.01(0.002) & 0.35(0.006) & 0.06(0.008) \\
    \hline
    \multicolumn{6}{c}{$n = 100, p = 200, p_{11} = 10, Horseshoe$} \\
    \hline
    PTG  & 0.30(0.015) & 0.24(0.014) & 0.08(0.016) & 0.37(0.016) & 0.38(0.019) \\
    GMM & 0.33(0.011) & 0.26(0.011) & 0.03(0.008) & 0.35(0.012) & 0.16(0.014) \\
    \hline
\end{tabular}}}
\captionof{table}{Empirical estimates of TPR and FDR in simulations under $n=100, p = 200$, $p_{11}$ is the number of true active mediators. The results are based on 200 replicates for each setting, and the standard errors are shown within parentheses. TPR(FDR=0.1) is the true positive rate controlled at a fixed FDR of 10\%; TPR(PIP$>$0.9) and FDR(PIP$>$0.9) are the empirical estimates when the PIP threshold for declaring active mediators is 0.9; TPR(PIP$>$0.5) and FDR(PIP$>$0.5) are the empirical estimates when the PIP threshold for declaring active mediators is 0.5.}
\label{tbl:efdr1}
\end{adjustwidth}
\end{table}

\begin{table}[!h]
\begin{adjustwidth}{0.0cm}{}
\scalebox{0.9}{
\centering{
  \begin{tabular}{c  c c c c c}
    \hline     
    Method & TPR(FDR=0.1) & TPR(PIP$>$0.9) & FDR(PIP$>$0.9) & TPR(PIP$>$0.5) & FDR(PIP$>$0.5)\\
    \hline
    \multicolumn{6}{c}{$n = 1000, p = 2000, p_{11} = 100, \textit{fixed effects (I)}$} \\
    \hline
    PTG  & 0.64(0.008) & 0.49(0.017) & 0.01(0.002) & 0.55(0.017) & 0.06(0.016) \\
    GMM & 0.61(0.009) & 0.40(0.004) & 0.01(0.003) & 0.55(0.005) & 0.07(0.010) \\
    \hline
    \multicolumn{6}{c}{$n = 1000, p = 2000, p_{11} = 100, \textit{fixed effects (II)}$} \\
    \hline
    PTG  & 0.40(0.008) & 0.20(0.004) & 0.01(0.003) & 0.37(0.012) & 0.07(0.010) \\
    GMM & 0.48(0.006) & 0.29(0.003) & 0.01(0.002) & 0.43(0.004) & 0.06(0.007) \\
    \hline
    \multicolumn{6}{c}{$n = 1000, p = 2000, p_{11} = 100, PTG$} \\
    \hline
    PTG & 0.40(0.008) & 0.19(0.004) & 0.01(0.011) & 0.44(0.007) & 0.13(0.006) \\
    GMM & 0.37(0.010) & 0.10(0.004) & 0.05(0.008) & 0.47(0.006) & 0.17(0.007) \\
	\hline
	\multicolumn{6}{c}{$n = 1000, p = 2000, p_{11} = 100, Gaussian$} \\
    \hline
    PTG & 0.42(0.006) & 0.20(0.005) & 0.03(0.002) & 0.51(0.005) & 0.17(0.004) \\
    GMM & 0.51(0.007) & 0.36(0.005) & 0.01(0.002) & 0.49(0.006) & 0.10(0.004) \\
    \hline
    \multicolumn{6}{c}{$n = 1000, p = 2000, p_{11} = 100, Horseshoe$} \\
    \hline
    PTG & 0.29(0.008) & 0.30(0.004) & 0.05(0.006) & 0.39(0.008) & 0.24(0.004) \\
    GMM & 0.38(0.007) & 0.35(0.004) & 0.03(0.003) & 0.45(0.004) & 0.18(0.015) \\
    \hline
\end{tabular}}}
\captionof{table}{Empirical estimates of TPR and FDR in simulations under $n=1000, p = 2000$, $p_{11}$ is the number of true active mediators. The results are based on 200 replicates for each setting, and the standard errors are shown within parentheses. TPR(FDR=0.1) is the true positive rate controlled at a fixed FDR of 10\%; TPR(PIP$>$0.9) and FDR(PIP$>$0.9) are the empirical estimates when the PIP threshold for declaring active mediators is 0.9; TPR(PIP$>$0.5) and FDR(PIP$>$0.5) are the empirical estimates when the PIP threshold for declaring active mediators is 0.5.}
\label{tbl:efdr2}
\end{adjustwidth}
\end{table}

We performed simulations on a single core of Intel(R) Xeon(R) Platinum 8176 CPU @ 2.10GHz, and the runtime comparison of the proposed methods is shown in Table \ref{table:runtime}. For both the small sample scenario with $n = 100$, $p = 200$, and the large sample scenario with $n = 1000$, $p = 2000$, the proposed algorithms can be finished in a reasonable amount of time. We still acknowledge that future development of new algorithms and/or new methods will likely be required to scale our method to handle thousands of individuals and millions of mediators. 

\begin{table}
\centering
{
  \begin{tabular}{c  c  c }
    \hline                  
	  Method & $n = 100$, $p = 200$ & $n = 1000$, $p = 2000$  \\
      \hline
      PTG & 30.5sec & 23.0min \\
      GMM & 88.8sec & 29.8min \\
      \hline
\end{tabular}}
\captionof{table}{The average runtime of the proposed methods for $n = 100$, $p = 200$ and $n = 1000$, $p = 2000$ in the simulations. Comparison was carried out on a single core of Intel(R) Xeon(R) Platinum 8176 CPU @ 2.10GHz. For the proposed methods, we in total ran 150,000 iterations.} 
\label{table:runtime}
\end{table}

\section{Detailed Description of MESA Data}
\hfill \break
MESA is a population-based longitudinal study designed to identify risk factors for the progression of subclinical cardiovascular disease (CVD) \citep{bild2002multi}. A total of 6,814 non-Hispanic white, African-American, Hispanic, and Chinese-American women and men aged 45$-$84 without clinically apparent CVD were recruited between July 2000 and August 2002 from the following 6 regions in the US: Forsyth County, NC; Northern Manhattan and the Bronx, NY; Baltimore City and Baltimore County, MD; St. Paul, MN; Chicago, IL; and Los Angeles County, CA. Each field center recruited from locally available sources, which included lists of residents, lists of dwellings, and telephone exchanges. Neighborhood socioeconomic disadvantage scores for each neighborhood were created based on a principal components analysis of 16 census-tract level variables from the 2000 US Census. These variables reflect dimensions of education, occupation, income and wealth, poverty, employment, and housing. For the neighborhood measures, we use the cumulative average of the measure across all available MESA examinations. The descriptive statistics for the exposure and outcome can be found in Table \ref{tbl:descriptive}.

In the MESA data, between April 2010 and February 2012 (corresponding to MESA Exam 5), DNAm were assessed on a random subsample of 1,264 non-Hispanic white, African-American, and Hispanic MESA participants aged 55$-$94 from the Baltimore, Forsyth County, New York, and St. Paul field centers. After excluding respondents with missing data on one or more variables, we had phenotype and DNAm data from purified monocytes on a total of 1,225 individuals and we focused on this set of individuals for analysis. The detailed description of DNAm data collection, quantitation and data processing procedures can be found in Liu et al \citep{liu2013methylomics}. Briefly, the Illumina HumanMethylation450 BeadChip was used to measure DNAm, and bead-level data were summarized in GenomeStudio. Quantile normalization was performed using the \textit{lumi} package with default settings \citep{du2008lumi}. Quality control (QC) measures included checks for sex and race/ethnicity mismatches and outlier identification by multidimensional scaling plots. Further probe filtering criteria included: ``detected'' DNAm levels in $<$90\% of MESA samples (detection $p$-value cut-off = 0.05), existence of a SNP within 10 base pairs of the target CpG site, overlap with a non-unique region, and suggestions by DMRcate \citep{chen2013discovery} (mostly cross-reactive probes). Those procedures leave us 403,713 autosomal probes for analysis. 

For computational reasons, we first selected a subset of CpG sites to be used in the final multivariate mediation analysis model. In particular, for each single CpG site in turn, we fit the following linear mixed model to test the marginal association between the CpG site and the exposure variable:
\begin{equation}
    M_i = A_i\psi_a + \boldsymbol{C_{1i}}^T\boldsymbol{\psi_c} + \boldsymbol{Z_i}^T\boldsymbol{\psi_u} + \epsilon_i, i = 1, ..., n
\end{equation}
where $A_i$ represents neighborhood SES value for the $i$'th individual and $\psi_a$ is its coefficient; $\boldsymbol{C_{1i}}$ is a vector of covariates that include age, gender, race/ethnicity, childhood socioeconomic status, adult socioeconomic status and enrichment scores for each of 4 major blood cell types (neutrophils, B cells, T cells and natural killer cells) to account for potential contamination by non-monocyte cell types; $\boldsymbol{Z_i}^T\boldsymbol{\psi_u}$ represent methylation chip and position random effects and are used to control for possible batch effects. The error term $\epsilon_i \sim MVN (0, \sigma^2 I_n)$ and is independent of the random effects. We obtained $p$-values for testing the null hypothesis $\psi_a=0$ from the above model. We further applied the R/Bioconductor package BACON \citep{van2017controlling} to these $p$-values to further adjust for possible inflation using an empirical null distribution. Based on these marginal $p$-values, we selected top 2,000 CpG sites with the smallest $p$-values for our Bayesian multivariate analysis.

Besides the proposed methods, we also implement the other competing methods on the MESA data. HIMA identifies one CpG site in the gene region of \textit{PCID2} as active mediator through the joint significance test with adjusted $p$-value = 6.3e-5, and this single site has also been detected by PTG and GMM methods. We apply the Pathway Lasso and bi-Lasso on multiple permuted data, and notice that same active mediators with non-zero indirect effects have been picked out in both original and permuted data. Thus, the signals identified by Pathway Lasso and bi-Lasso are very probably false discoveries. For BAMA, the estimated PIPs over the 2,000 CpG sites are no more than 0.1, which does not provide strong evidence on the finding of active mediators.

\begin{table}[!h]
\centering{
  \begin{tabular}{l c c c}
    \hline                  
	& \begin{tabular}[x]{@{}c@{}}\textbf{Full}\\\textbf{Sample}\\\textbf{(n, \%)}\end{tabular} &  \begin{tabular}[x]{@{}c@{}}\textbf{Neighborhood}\\\textbf{Socioeconomic}\\\textbf{Disadvantage}\\\textbf{Mean (SD)}\end{tabular} & \begin{tabular}[x]{@{}c@{}}\textbf{Body Mass Index (BMI)}\\\textbf{Mean (SD)}\end{tabular} \\
	\hline
    \textbf{Full sample} & 1225 (100) & -0.32 (1.11) & 29.5 (5.49) \\
    \textbf{Age} & & & \\
    \multicolumn{1}{r}{55$-$65 years} & 462 (38) & -0.18 (0.96) & 30.3 (6.02)   \\
    \multicolumn{1}{r}{66$-$75 years} & 397 (32) & -0.30 (1.16) & 30.1 (5.21) \\
    \multicolumn{1}{r}{76$-$85 years} & 300 (24) & -0.47 (1.15) & 28.2 (4.65)\\
    \multicolumn{1}{r}{86$-$95 years} & 66 (5) & -0.67 (1.46) & 26.6 (4.66) \\
    \textbf{Race/ethnic group} & & & \\
    \multicolumn{1}{r}{Non-Hispanic white} & 580 (47) & -0.56 (1.18) & 28.7 (5.40) \\
    \multicolumn{1}{r}{African-American} & 263 (22) & -0.16 (0.98) & 30.5 (5.69) \\
    \multicolumn{1}{r}{Hispanic} & 382 (31) & -0.05 (1.00) & 30.0 (5.32) \\
    \textbf{Gender} & & & \\
    \multicolumn{1}{r}{Female} & 633 (52) & -0.24 (1.09) & 30.1 (6.20) \\
    \multicolumn{1}{r}{Male} & 592 (48) & -0.40 (1.12) & 28.9 (4.54)  \\
    \hline
\end{tabular}}
\captionof{table}{Characteristics of 1225 participants from MESA. \%: proportion in the corresponding category. SD: standard deviation. }
\label{tbl:descriptive}
\end{table}

\section{Detailed Description of LIFECODES Data}
\hfill \break
The LIFECODES prospective birth cohort enrolled approximately 1,600 pregnant women between 2006 and 2008 at the Brigham and Women’s Hospital in Boston, MA. Participants between
20 and 46 years of age were all at less than 15 weeks gestation at the initial study visit, and followed up to four visits (targeted at median 10, 18, 26, and 35 weeks gestation). At the initial study visit, questionnaires were administered to collect demographic and health-related information. Subjects' urine and plasma samples were collected at each study visit. Among participants recruited in the LIFECODES cohort, 1,181 participants were followed until delivery and had live singleton infants. The birth outcome, gestational age, was also recorded at delivery, and preterm birth was defined as delivery prior to 37 weeks gestation. This study received institutional review board (IRB) approval from the Brigham and Women’s Hospital and all participants provided written informed consent. All of the methods within this study were performed in accordance with the relevant guidelines and regulations approved by the IRB. Additional details regarding recruitment and study design can be found in \citep{mcelrath2012longitudinal, ferguson2014variability}. 

In this study, we focused on a subset of $n = 161$ individuals with their urine and plasma samples collected at one study visit occurring between 23.1 and 28.9 weeks gestation (median = 26 weeks). Characteristics of the subset sample is described in Table \ref{tbl:descriptive1}. Subjects' urine samples were refrigerated ($4^\circ$C) for a maximum of 2 hours before being processed and stored at $-80^\circ$C. Approximately 10mL of blood was collected using ethylenediaminetetraacetic acid plasma tubes and temporarily stored at $4^\circ$C for less than 4 hours. Afterwards, blood was centrifuged for 20 minutes and stored at $-80^\circ$C. Environmental exposure analytes were measured from urine samples by NSF International in Ann Arbor, MI, following the methods developed by the Centers for Disease Control (CDC) \citep{silva2007quantification}. Those exposure analytes include phthalates, phenols and parabens, trace metals and polycyclic aromatic hydrocarbons. To adjust for urinary dilution, specific gravity (SG) levels were measured in each urine sample using a digital handheld refractometer (ATAGO Company Ltd., Tokyo, Japan), and was included as a covariate in regression models. Urine and plasma were subsequently analyzed for endogenous biomarkers, including 51 eicosanoids, five oxidative stress biomarkers and five immunological biomarkers in the present study. For a detailed description of the biomarkers that we analyzed and the media (urine or plasma) in which they were measured, please refer to \citep{aung2019prediction}.

\begin{table}[!h]
\centering{
  \begin{tabular}{l c c c}
    \hline                  
	& \begin{tabular}[x]{@{}c@{}}\textbf{Full}\\\textbf{Sample}\\\textbf{(n = 161)}\end{tabular} &  \begin{tabular}[x]{@{}c@{}}\textbf{Preterm}\\\textbf{(<37 weeks gestation,}\\\textbf{n = 52)}\end{tabular} & \begin{tabular}[x]{@{}c@{}}\textbf{Control}\\\textbf{(n = 109)}\end{tabular} \\
	\hline
    \textbf{Age$^\mathrm{a}$} & 32.7 (4.4) & 32.1 (5.0) & 33.0 (4.2) \\
    \textbf{BMI at Initial Visit$^\mathrm{a}$} & 26.7 (6.4) & 28.5 (7.6) & 25.8 (5.6) \\
    \textbf{Race/ethnic group$^\mathrm{b}$} & & & \\
    \multicolumn{1}{r}{White} & 102 (63\%) & 29 (56\%) & 73 (67\%) \\
    \multicolumn{1}{r}{African-American} & 18 (11\%) & 7 (13\%) & 11 (10\%)\\
    \multicolumn{1}{r}{Other} & 41 (26\%) & 16 (31\%) & 25 (23\%) \\
    \textbf{Gestational weeks$^\mathrm{a}$} & 37.5 (3.1) & 34.1 (3.2) & 39.1 (1.1)\\
    \hline
\end{tabular}}
\captionof{table}{Characteristics of all participants in the subset sample from the LIFECODES prospective birth cohort
(n = 161). $^\mathrm{a}$Continuous variables presented as: mean (standard deviation). $^\mathrm{b}$Categorical variables
presented as: count (proportion).}
\label{tbl:descriptive1}
\end{table}

\bibliographystyle{rss}
\bibliography{reference}